\documentclass{aa}

\usepackage{amsmath,amssymb,graphicx,txfonts,epsf,natbib}
\bibpunct{(}{)}{;}{(a)}{}{,}

\newcommand{\EE}{{\mathbb E}}
\newcommand{\PP}{{\mathbb P}}
\newcommand{\RR}{{\mathbb R}}

\newcommand{\cL}{{\cal L}}
\newcommand{\cN}{{\cal N}}
\newcommand{\cS}{{\cal S}}

\newcommand{\1}{{\bf 1}}

\begin{document}

\title{Order statistics and heavy-tail distributions for planetary
perturbations on Oort cloud comets}

\titlerunning{Order statistics and heavy-tail distributions}

\author{R. S. Stoica\inst{1}\and S. Liu \inst{1}\and Yu. Davydov\inst{1}
\and M. Fouchard \inst{2,3}\thanks{\emph{Present address:} Observatoire de Lille, 1 Impasse de l'Observatoire, 59\,000 Lille, France.}\and A. Vienne \inst{2,3}\and G.B. Valsecchi \inst{4}}
\offprints{Marc Fouchard, \\
\email{marc.fouchard@univ-lille1.fr}}

\authorrunning{Stoica et al.}

\institute{
University Lille 1, Laboratoire Paul Painlev\'e, 59655 Villeneuve d'Ascq Cedex, France 
\and University Lille 1, LAL, 59000 Lille, France
\and Institut de M\'ecanique C\'eleste et Calcul d'Eph\'em\'erides, 77 av. Denfert-Rochereau,
75014 Paris, France
\and INAF-IASF, via Fosso del Cavaliere 100, 00133 Roma, Italy}

\date{Received/Accepted}

\abstract{}
{This paper tackles important aspects of comets dynamics from a
statistical point of view. Existing methodology uses numerical
integration for computing planetary perturbations for simulating such
dynamics. This operation is highly computational. It is reasonable to
wonder whenever statistical simulation of the perturbations can be much
more easy to handle.} 
{The first step for answering such a question is
to provide a statistical study of these perturbations in order to
catch their main features. The statistical tools used are order
statistics and heavy tail distributions.} 
{The study carried out
indicated a general pattern exhibited by the perturbations around the
orbits of the important planet. These characteristics were validated
through statistical testing and a theoretical study based on \"Opik
theory.}{} 

\maketitle

\section{Introduction}
Comet dynamics is one of the most difficult phenomena to
model in celestial mechanics.  Indeed their dynamics is strongly
chaotic, thus individual motions of known comets are hardly
reproducible for more than a few orbital periods.  When the origin of comets
is under investigation, one is thus constrained to make use of
statistical tools in order to model the motion of a huge number of
comets supposed to be representative of the actual population. Such
statistical model should also be reliable on a time scale comparable to the 
age of the solar system.

Due to their very elongated shapes, comet trajectories are affected by
planetary perturbations during close encounters with planets.  Such
perturbations turn out to be the main mechanisms able to affect comet
trajectories. Consequently, it is of major importance to model these
perturbations in a way which is statistically reliable and with the
lowest cost in computing time.

A direct numerical integration of a 6 bodies restricted problem (Sun,
Jupiter, Saturn, Uranus, Neptune, Comet) each time a comet enters the
planetary region of the Solar System is not possible due to the cost
in computer time.

Looking for an alternative approach, we can take advantage of the fact
that planetary perturbation on Oort cloud comets are uncorrelated. In
fact the orbital period of such comets are so much larger than those
of the planets, that when the comet returns, the phases of the latter
can be taken at random.  Thus we can build a synthetic integrator \`a la
Froeschl\'e and Rickman \citep{FRO.RIC:81} to speed up the
modeling. The criticism by \citep{FOUetal:03} to such an approach does
not apply in the present case because, as just said, successive
planetary perturbations on an Oort cloud comets are uncorrelated.

The aim of this paper is to give a statistical description of a large
set of planetary perturbations assumed to be representative of those
acting on Oort cloud comets entering the planetary region. To this
purpose we use order statistics and heavy tails distributions.

The rest of this paper is organised as follow.  Section 2 is devoted
to the presentation of the mechanism producing the data, {\it i.e.}
the planetary perturbations and the statistical tools used to analyse
the data. These tools are order statistics and heavy-tail
distributions, that allow, respectively, the study and the modeling of
the data distribution, with respect to its symmetry, skewness and tail
fatness. The obtained results are shown and interpreted in the third
section. The results are finally analysed from a more theoretical
point of view using the \"Opik theory in Section 4. The paper closes
with conclusions and perspectives.

\section{Statistical tools}
\subsection{Data compilation}
By planetary perturbations, one intends the variations of the orbital
parameters between their values before entering the planetary region
of the Solar System, {\it i.e.} the barycentric orbital element of the
osculating cometary orbit $(z_i,q_i,\cos i_i,\omega_i,\Omega_i)^T$
(where $q$, $i$, $\omega$, $\Omega$ are the perihelion distance, the
inclination, the argument of perihelion and the longitude of the
ascending node and $z=-1/a$ with $a$ the semi-major axis), and their
final values $(z_f,q_f,\cos i_f,\omega_f,\Omega_f)^T$, that is either
when the comet is at its aphelion or when it is back on a keplerian
barycentric orbit.

Between its initial and final values, the system Sun + Jupiter +
Saturn + Uranus + Neptune + comet is integrated using the RADAU
integrator at the 15th order \citep{EVE:85} for a maximum of
2\,000~yrs.  Then the planetary perturbation obtained through this
integration is $(\Delta z = z_f-z_i, \Delta q = q_f-q_i,\Delta \cos i
= \cos i_f-\cos i_i, \Delta \omega = \omega_f-\omega_i, \Delta \Omega
= \Omega_f-\Omega_i)^T$.  The detail on the numerical experiment used
to perform the integrations may be found in \cite{RICetal:01}.

Repeating the above experiment with a huge number of comets (namely
$9\,600\,000$), one gets a set of planetary perturbations. The comets
are chosen with uniform distribution of the perihelion distance
between 0 and 32 AU, cosine of the ecliptic inclination between -1 and
1 and argument of perihelion, longitude of the ascending node between
0 and $360^\circ$. The initial mean anomaly is chosen such that the
perihelion passage on its initial keplerian orbit occurs randomly with
an uniform distribution between 500 and 1\,500 years after the
beginning of the integration.

In the present study, because the perturbations are mainly depending
on $q_i$ and $\cos i_i$ \citep{FER:81}, each perturbation is
associated to the couple $(\cos i_i,q_i)$. Similarly, since the
orbital energy is the main quantity which is affected by the planetary
perturbations, we will consider only these perturbations here.

Consequently, our data are composed by a set of triplets $(\cos i_i,
q_i, Z)$ where $Z=z_f-z_i$ denotes the perturbations of the cometary
orbital energy by the planets, and $(\cos i_i, q_i)$ a point in a
space denoted by $K$.  In the following, we call $Z$ the perturbation
mark.

\subsection{Exploratory analysis based on order statistics}
Let $Z_1,\ldots,Z_n$ be a sequence of independent identically
distributed random variables and let $F(z)=P(Z \leq z), z \in \RR$ be
the corresponding cumulative distribution function. Let us consider
also $\Sigma_n$, the set of permutations on $\{1,\ldots,n\}$.

The order statistics of the sample $(Z_1,\ldots,Z_n)$ is the
rearrangement of the sample in increasing order and it is denoted by
$(Z_{(1,n)},\ldots,Z_{(n,n)})$. Hence $Z_{(1,n)} \leq \ldots,\leq
Z_{(n,n)}$ and there exists a random permutation $\sigma_n \in \Sigma_n$
such that 
\begin{equation}
(Z_{(1,n)},\ldots,Z_{(n,n)})=(Z_{\sigma_{n}(1)},\ldots,Z_{\sigma_{n}(n)}).
\label{def_orderstat}
\end{equation}

In the following, some classical results from the literature are
presented~\citep{DAV:81,DELetal:06}. If $F$ is continuous, then almost surely
$Z_{(1,n)} < \ldots,< Z_{(n,n)}$ and the permutation $\sigma_n$ in
definition~(\ref{def_orderstat}) is unique. If $Z_1$ has a probability
density $f$, then the probability density of the order statistics is
given by
\begin{equation*}
n!\1\{z_1 < \ldots z_n\}f(z_1)\ldots f(z_n).
\end{equation*}

A major characteristic of order statistics is that they allow quantiles
approximations. The quantiles are one of the most easy to use tool
for characterising a probability distribution. In
practice, the data distribution can be described by such empirical
quantiles.

Two important results are now presented. The first result shows how to
compute empirical quantiles using order statistics. Let us assume that
$F$ is continuous and there exists an unique solution $z_q$ to the
equation $F(z) = q$ with $q \in (0,1)$. Clearly, $z_q$ is the $q-$quantile of
$F$. Let $(k(n),n \geq 1)$ be an integers sequence such that $1 \geq
k(n) \geq n$ and $\lim_{n \rightarrow \infty} \frac{k(n)}{n} =
q$. Then the sequence of the empirical quantiles $(Z_{(k(n),n)},n \geq
1)$ converges almost surely towards $z_q$.

The second result allows the computation of confidence intervals and
hypothesis testing. If $Z_1$ has a continuous probability density
$f$ such that $f(z_q) > 0$ for $q \in (0,1)$ and if it is supposed
that $k(n) = nq +o(\sqrt{n})$, then $Z_{(k(n),n)}$ converges in
distribution towards $z_q$ as it follows
\begin{equation*}
\sqrt{n}(Z_{(k(n),n)}-z_q) \stackrel{\cL}{\rightarrow}\cN
\left(0,\frac{q(1-q)}{f(z_q)^2} \right).
\end{equation*}

The exploratory analysis we propose for the perturbation data sets is
based on the computation of empirical quantiles. There are several
reasons motivating such a choice. First, there is not too much a
priori knowledge concerning the perturbations marks, except that they
are distributed around zero and that they are uniformly located in
$K$. This implies that very few hypothesis with respect to the data
can be done. Clearly, in order to apply such an analysis the only
assumptions needed are the conditions of validity for the central
limit theorem. From a practical point of view, an empirical quantiles
based analysis allows for checking the tails, the symmetry and the
general spatial pattern of the data distribution. From a theoretical
point of view, the mathematics behind this tool allow a rather
rigorous analysis.

\subsection{Stable distributions models}
Stable laws are a rich class of probability distributions that allow
heavy tails, skewness and have many nice mathematical properties. They
are also known in the literature under the name of $\alpha$-stable,
stable Paretian or L\'evy stable distributions. These models were
introduced by~\cite{LEV:25}. In the following some basic notions and
results on stable distributions are given~\citep{BORetal:05,FEL:71,SAMetal:94}.

A random variable $Z$ has a {\em stable distribution} if for any
$A,B>0$, there is a $C>0$ and $D\in\RR^1$ such that
\begin{equation*}
\label{def1}
AZ_{1}+BZ_{2}{\stackrel{\cL}{=}}CZ+D,
\end{equation*} 
where $Z_{1}$ and $Z_{2}$ are independent copies of $Z$, and
"${\stackrel{\cL}{=}}$" denotes equality in distribution.

\begin{figure*}[!htbp]
\begin{center}
\begin{tabular}{cc}
a)\epsfxsize=5cm \epsffile{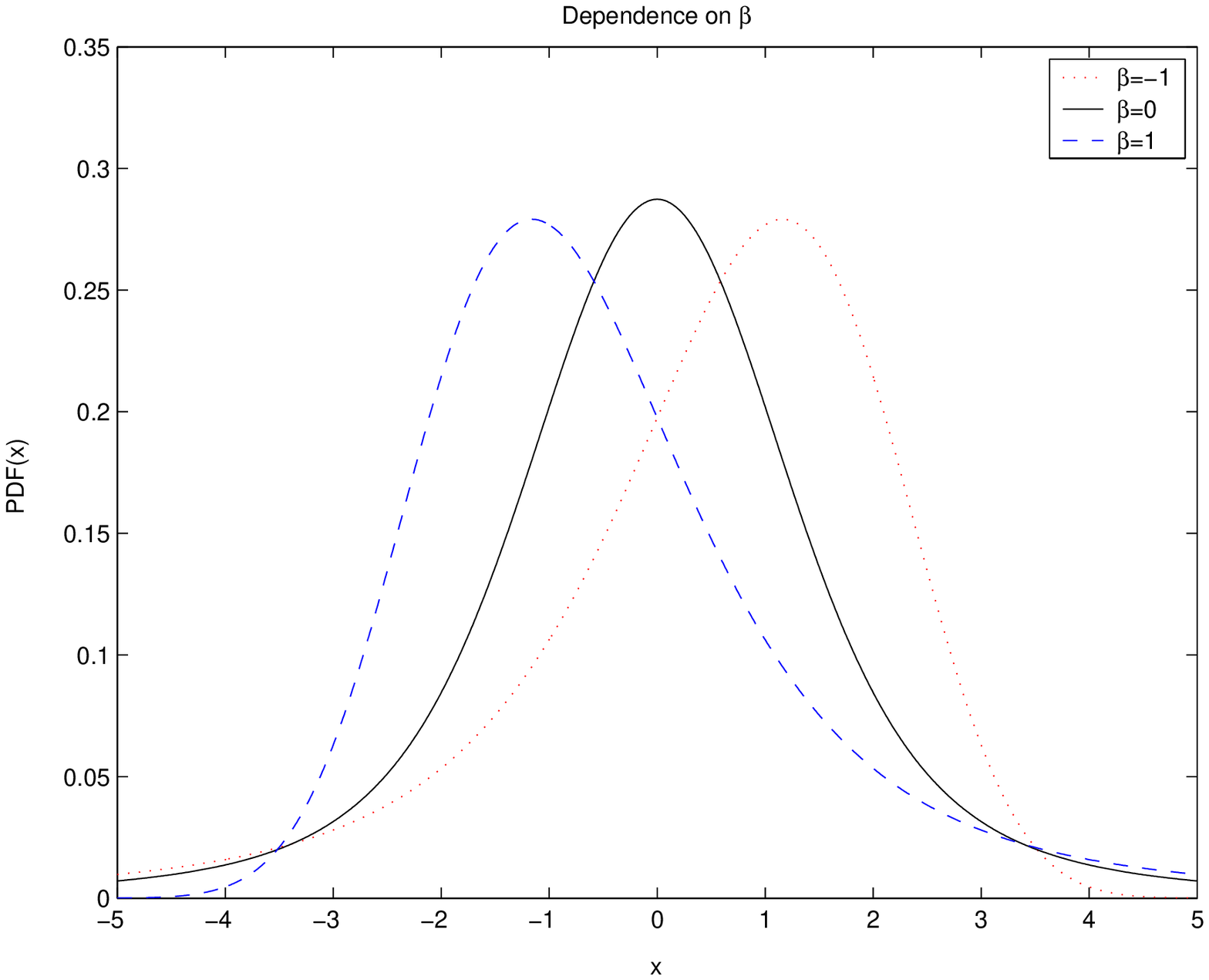} & 
b)\epsfxsize=5cm \epsffile{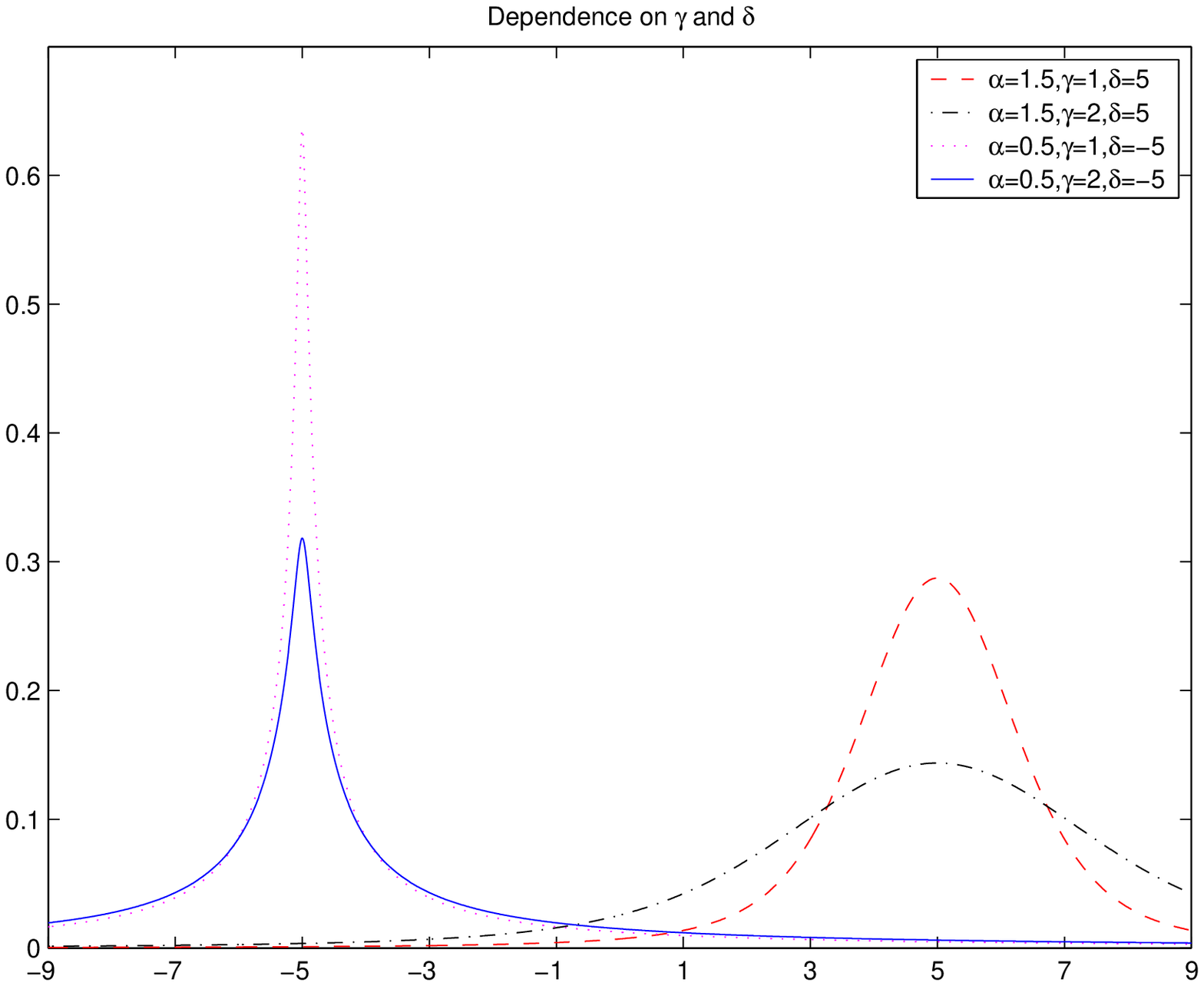}\\
\end{tabular}
\end{center}
\caption{Influence of the parameters on the shape of a stable
distribution~: a) $\beta$ parameter, b) $\alpha,\gamma$ and $\delta$
parameters.}
\label{stable_parameters}
\end{figure*}

A stable distribution is characterised by four parameters $\alpha\in
(0,2]$, $\beta\in [-1,1]$, $\gamma\geqslant 0$ and $\delta
\in \mathbb{R}^1$ and it is denoted by
$\cS_{\alpha}(\beta,\gamma,\delta)$. The role of each parameter is as
it follows~: $\alpha$ determines the rate at which the distribution
tail converges to zero, $\beta$ controls the skewness of the
distribution, whereas $\gamma$ and $\delta$ are the scale and shift
parameters, respectively. Figure~\ref{stable_parameters} shows the
influence of these parameters on the distribution shape.

The linear transformation of stable random variable is also a stable
variable. If $\alpha \in (0,2)$, then $\EE |Z|^p<\infty$ for any
$0<p<\alpha$ and $\EE |Z|^p=\infty$ for any $p\geqslant\alpha$. The
distribution is Gaussian if $\alpha=2$. The stable variable with
$\alpha<2$ has an infinite variance and the corresponding distribution
tails are asymptotically equivalent to a Pareto
law~\citep{SKO:61}. More precisely
\begin{equation}\label{powerlaw}
\left\{\begin{array}{lll}\lim_{z\rightarrow\infty}z^\alpha \PP\{Z>z\}&=&\frac{(1+\beta)}{2}\sigma,\\
\lim_{z\rightarrow\infty}z^\alpha \PP\{Z<-z\}&=&\frac{(1-\beta)}{2}\sigma.\end{array}\right.
\end{equation}
where $\sigma=C_{\alpha}\gamma^\alpha$,
$C_{\alpha}=\frac{1-\alpha}{2\Gamma(2-\alpha)\cos(\pi\alpha/2)}$ if
$\alpha\neq 1$ and $C_{\alpha}=\frac{2}{\pi}$ elsewhere. The
distribution is symmetric whenever $\beta = 0$, or skewed
otherwise. In the case $\alpha<1$, the support of the distribution
$\cS_{\alpha}(\beta,\gamma,0)$ is the positive half-line when
$\beta=1$ and the negative half-line when $\beta=-1$. If $\alpha>1$,
then the first order moment exists and equals the shift parameter
$\delta$.

One of the technical difficulty in the study of stable distribution is that
except for a few cases (Gaussian, Cauchy and L\'evy), there is no
explicit form for the densities. The characteristic function can be
used instead, in order to describe the distribution. There exist
numerical methods able to approximate the probability density and the
cumulative distribution functions~\citep{NOL:97}. Simulation algorithms
for sampling stable distribution can be found
in~\cite{BORetal:05,CHAetal:76}.

Due to the previous considerations, parameter estimation is still an
open and challenging problem. Several methods are available in the
literature~\citep{FAMetal:71,MCC:86,MITetal:99,NOL:01,PRE:72}. Nevertheless,
these methods have all the same drawback, in the sense that the data
is supposed to be a sample of a stable law. It is a well known fact,
that if the data comes from a different distribution, the inference of
the tail index may be strongly misleading. A solution to this problem
is to estimate the tail exponent~\citep{HIL:75} and then estimate
distribution parameters if $\alpha \in (0,2]$. 

Still, it remains to solve the problem of parameter estimation
whenever the tail exponent is greater than $2$. Under these
circumstances, distributions with regularly varying tails can be
considered. A random variable has a distribution with regularly
varying tails of index $\alpha \geq 0$ if there exist $p,q \geq 0, p+q
= 1$ and a slowly varying function $L$, {\it i.e} $\lim_{z \rightarrow
\infty} \frac{L(\lambda z)}{L(z)} = 1$ for any $\lambda >0$, such that
\begin{equation}
\label{rvlaw}
\left\{\begin{array}{lll}\lim_{z\rightarrow\infty}z^\alpha L(z)\PP\{Z>z\}&=&p,\\
\lim_{z\rightarrow\infty}z^\alpha L(z)\PP\{Z<-z\}&=&q.\end{array}\right.
\end{equation}

It is important to notice that the conditions~(\ref{powerlaw})
can be obtained from~(\ref{rvlaw}) whenever $L(z) = 1/\sigma$ and
$p=(1+\beta)/2$. 

The parameter estimation algorithm proposed
by~\cite{DAVetal:99,DAVetal:04} is constructed under the assumption
that the sample distribution has the asymptotic
property~(\ref{powerlaw}). The algorithm gives three estimated values
$\widehat{\alpha}, \widehat{\beta}, \widehat{\sigma}$. The
$\widehat{\delta}$ can be computed easily whenever $\alpha > 1$, by
approximating it using the empirical mean of the samples. This
parameter estimation method can be used for stable distribution and in
this case, $\widehat{\alpha}$ should indicate positive values lower
than $2$. In the same time, the strong point of the method is that it
can be used for data not following stable distributions. In this case
the data distribution is assumed to have regularly varying tails. The
weak point of this algorithm is that in this case, it does not give
indications concerning the body of the distribution. Nevertheless, in
both cases, this method allows a rather complete characterisation of a
wide panel of probability distributions. The code implementing the
algorithm is available just by simple demand to the authors.

\section{Results}
\subsection{Empirical quantiles}
The lack of stationarity of the perturbations marks imposes the
partitioning of the location space in a finite number cells. Let us
consider such a partition $K=\cup_{i=1}^{n}K_i$. The cells $K_i$ are
disjoint and they all have the same volume. The size of volume has to
be big enough in order to contain a sufficient number of
perturbations. In the same time, the volume has to be small enough to
allow stationarity assumptions for the perturbations marks inside a
cell. After several trial and errors, we have opted for a partition
made of square cells $K_i$, having all the same volume $0.1 \times 0.1~$AU, 
so that each cell contains about $1\,500$ perturbations.

We were interested in three questions concerning the perturbations
marks distributions. The first two questions are related to the tails
and the symmetry of the data distribution. The third question is
related to a more delicate problem. It is a well known fact that the
perturbations locations follow an uniform distribution in
$K$. Nevertheless, much few is known about the spatial distribution of
the perturbations marks, except that they are highly dependent on
their corresponding locations.  So, the third question to be
formulated is the following~: do the distributions of the perturbations
marks exhibit any pattern depending on the perturbation location ?

For this purpose, empirical $q-$quantiles were computed in each
cell. The most part of these values were indicating that the
perturbations marks are distributed around the origin, while no
particular spatial pattern is exhibited in the perturbation location
space.

On the other hand, the situation is completely different for extremal
$q-$values such as~: $0.01,0.05,0.95,0.99$. These quantiles were
indicating rather important values around the semi-major axis of each
planet. In order to check if these values may reveal heavy tail
distributions, the difference based indicator $z_{q} -
\widehat{n}_{q}$ was built. The first term of this indicator
represents an empirical $q-$quantile. The second term is the
theoretical $q-$quantile of the normal law with mean and standard
deviation given by $z_{0.50}$ and $0.5(z_{0.84}-z_{0.16})$. Hence, for
values of $q$ approaching $1$, positive values of the indicator may
suggest heavy-tail behaviour for the data. Clearly, this indicator may
be used also for quantiles approaching $0$. In this case, it is the
negative sign that reveals the fatness of the distribution tail.

\begin{figure*}[!htbp]
\begin{center}
\begin{tabular}{cc}
a)\epsfxsize=5.5cm \epsffile{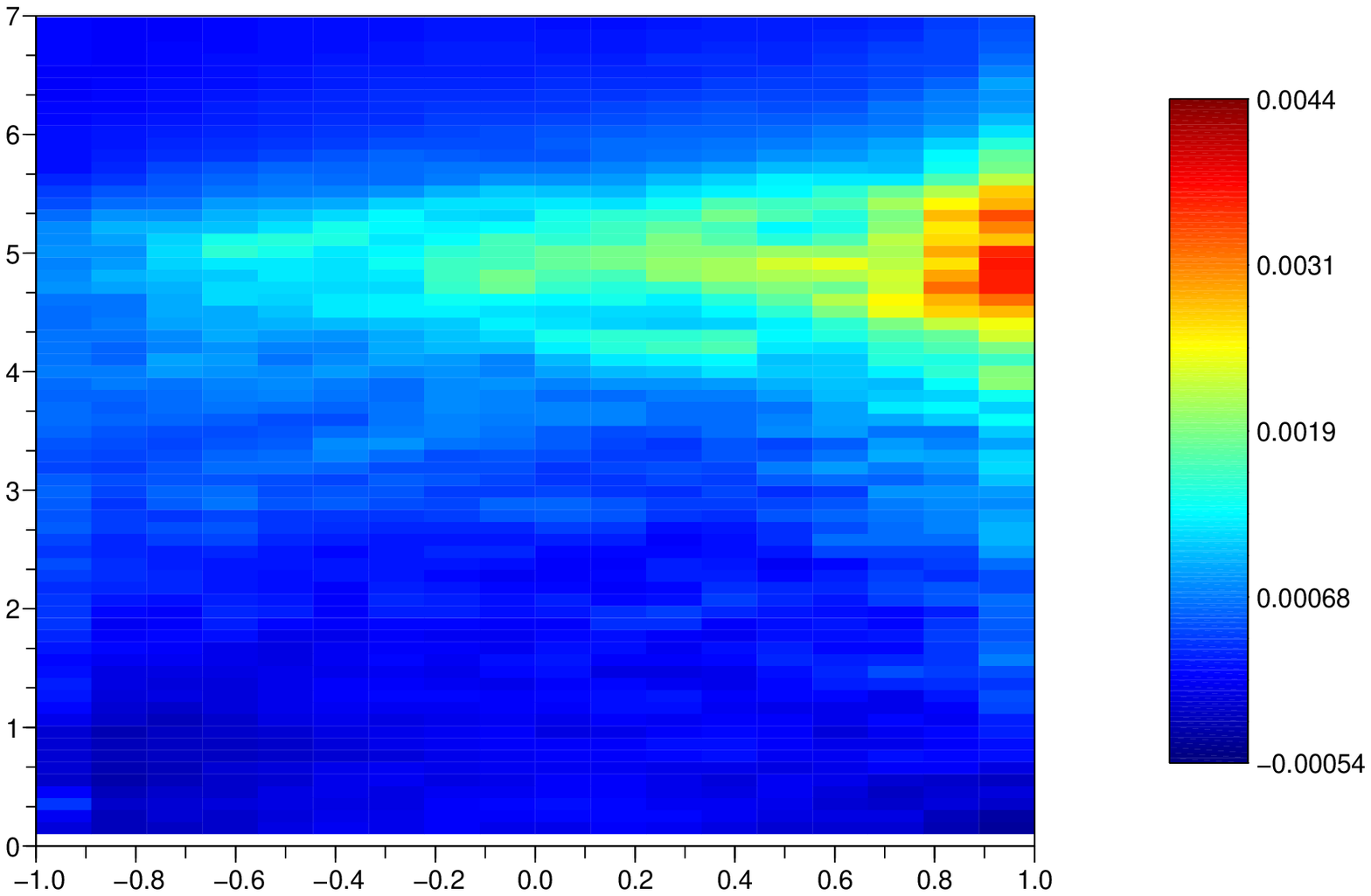} & 
b)\epsfxsize=5.5cm \epsffile{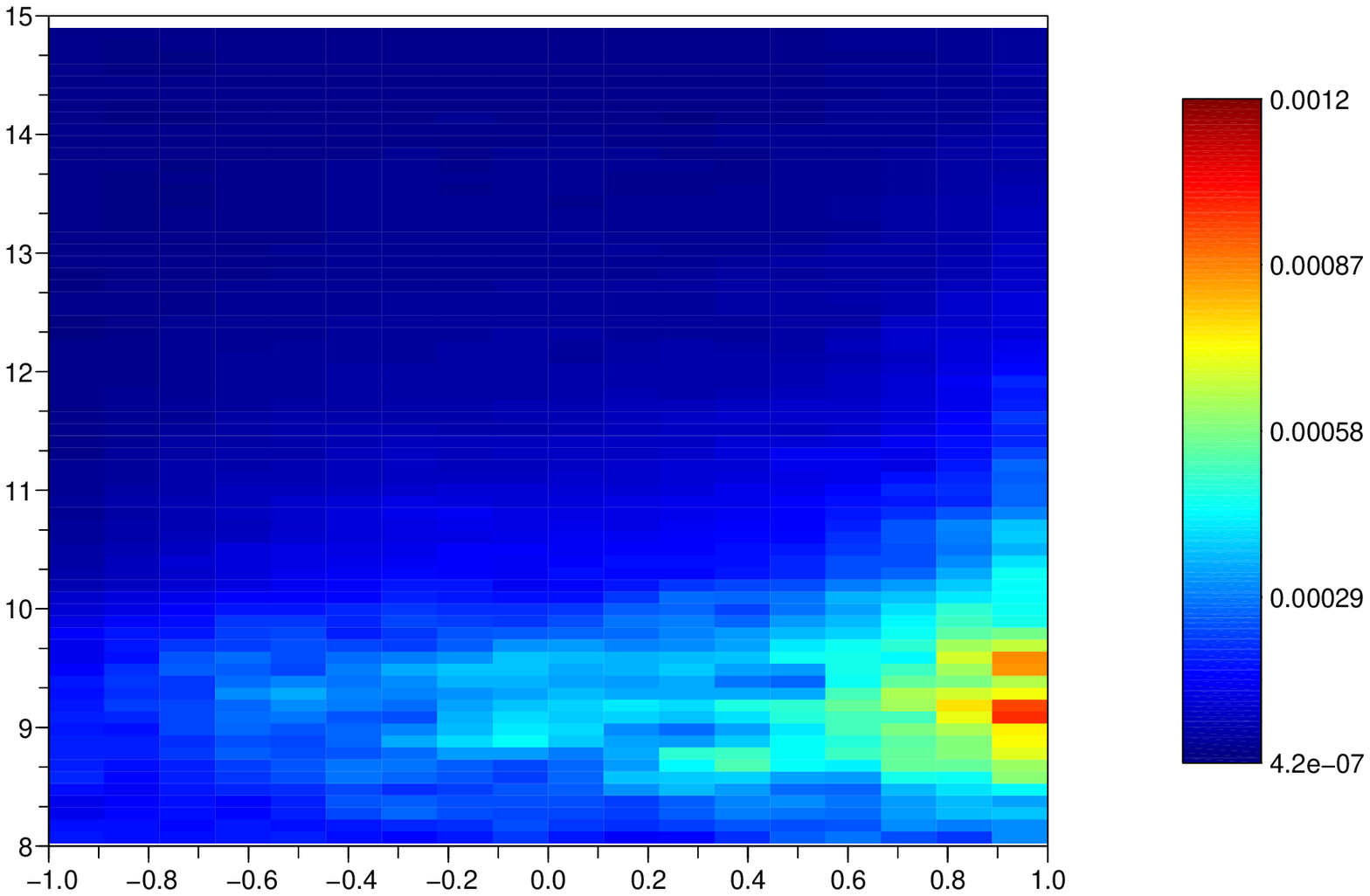}\\
c)\epsfxsize=5.5cm \epsffile{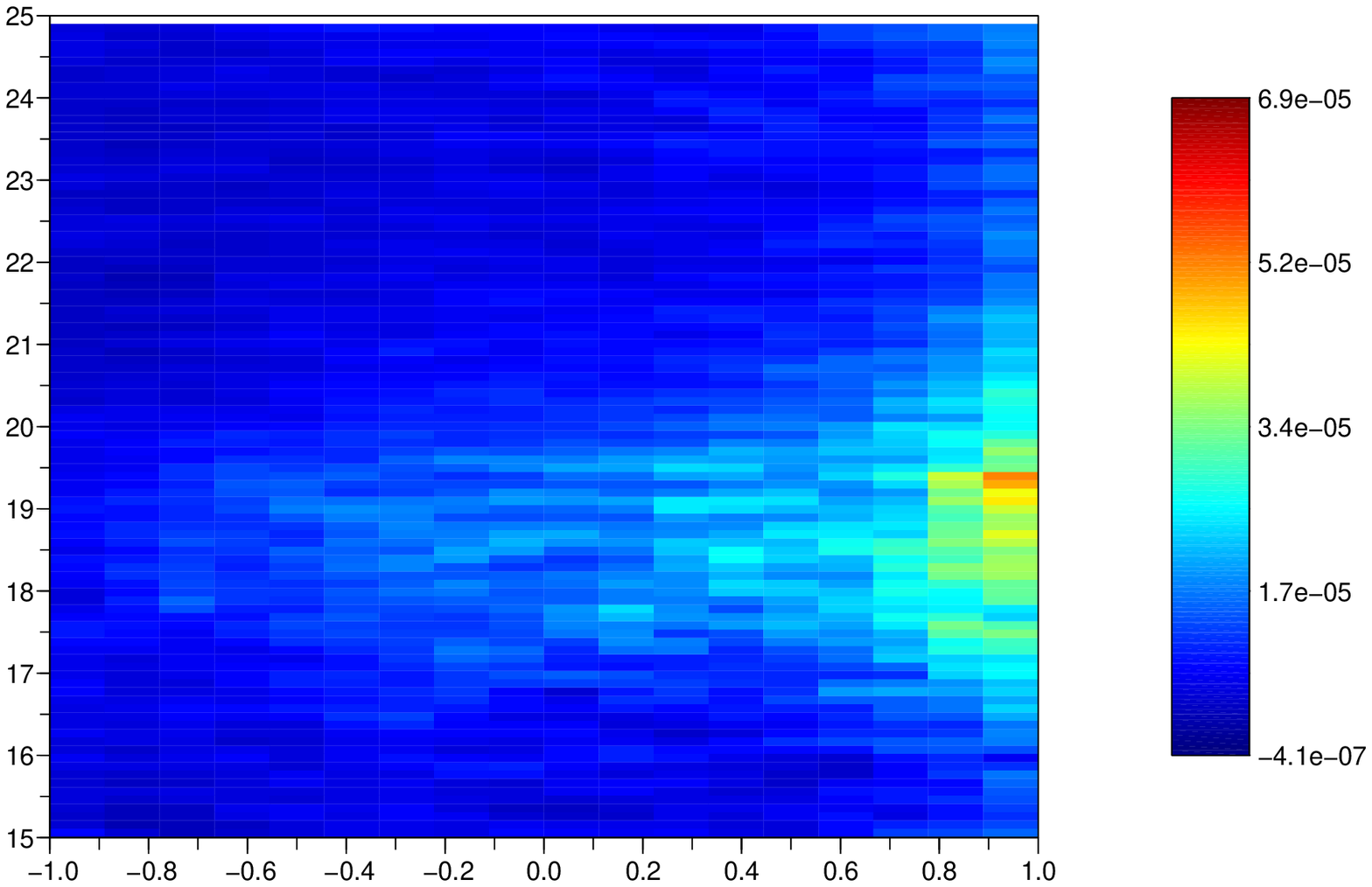} & 
d)\epsfxsize=5.5cm \epsffile{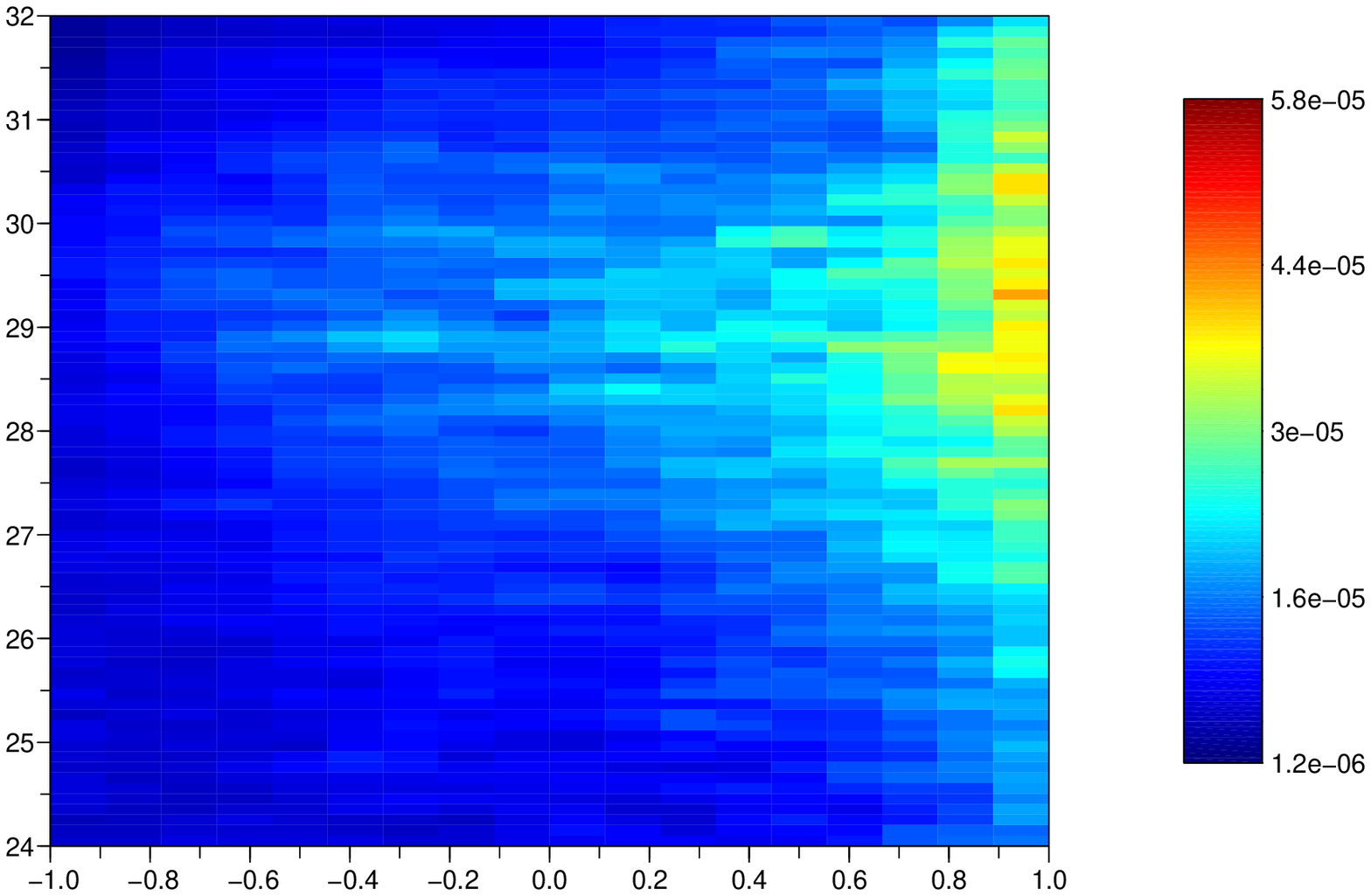}\\ 
\end{tabular}
\end{center}
\caption{ Empirical quantiles based difference indicator $z_{0.99} -
\widehat{n}_{0.99}$ for the perturbation marks around the big
planets~: a) Jupiter, b) Saturn, c) Uranus d) Neptune. For each
diagram the $y$-axis correspond to initial perihelion distance in AU,
and the $x$-axis to cosine of the inclination. We recall that the
respective semi-major axis of the four giant planets are:
$a_J=5.2$~AU, $a_S=9.6$~AU, $a_U=19.2$~AU, $a_N=30.1$~AU.}
\label{exploratory_tails}
\end{figure*}

In Figure~\ref{exploratory_tails} the values obtained for the
difference indicator $z_{0.99}-\widehat{n}_{0.99}$ are shown. It can
be observed that its rather important values appeared whenever the
perturbations are located in the vicinity of a planet orbit. All these
values tend to form a spatial pattern similar to an arrow-like
shape. As it can be observed, this shape is situated around the planet
orbit and it is pointing from the right to left. It tends to vanish,
while the cosine of the inclination angle approaches $-1$. The
prominence of this arrow shape clearly depends on the closest planet~:
bigger the planet is, sharper is the arrow-like shape. This can be
observed by looking at the change of values for the
difference indicator with respect the size of the planet. These
observations fulfil some good sense expectations~: the comets
perturbations tend to be more important whenever a comet cross the
orbit of a giant planet.

Since these phenomena are observed for extremal $q-$quantiles,
they indicate that the distribution tails may be an important feature
for the data. Hence, a statistical model for the data should be able to
catch these characteristics of the perturbations marks.

\begin{figure}[!htbp]
\begin{center}
\begin{tabular}{c}
a)\epsfxsize=5.5cm \epsffile{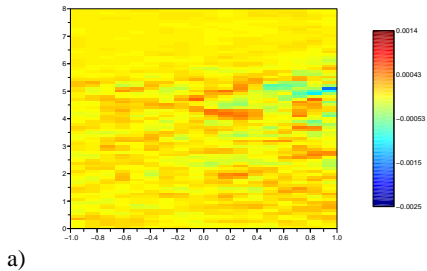}
\end{tabular}
\end{center}
\caption{Exploring symmetry using empirical quantiles difference $z_{0.99} - |z_{0.01}|$ 
for the perturbations marks around Jupiter. Axis are as for Fig.~\ref{exploratory_tails}}
\label{exploratory_simmetry}
\end{figure}

Empirical quantiles can be also used in straightforward way as
symmetry indicators of the data distribution. Clearly, by just
checking whenever the difference $z_q - |z_{1-q}|$ tends to $0$, this
may suggest a rather symmetric data
distribution. Figure~\ref{exploratory_simmetry} shows the computation
of such differences for each data cell. The values obtained are rather
small all over the studied region. Nevertheless, there are some
regions and especially around the Jupiter's orbit we may suspect the
data distributions are a little bit skewed. Still, since the
perturbations have rather small numerical values, assessing symmetry
using the proposed indicator has to be done cautiously.

It is reasonable to expect a more reliable answer concerning this
question by using a statistical model. Clearly, such a model should be able
to catch the symmetry of the data distribution as well.

\begin{figure}[!htbp]
\begin{center}
\begin{tabular}{c}
\epsfxsize=5.5cm \epsffile{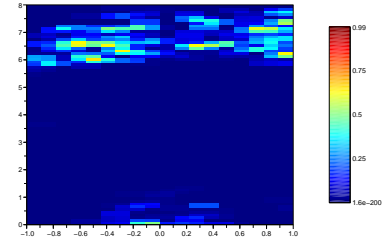}
\end{tabular}
\end{center}
\caption{$p-$values computed for testing the normality of the empirical 
quantiles $z_{0.95}$ around Jupiter.}
\label{exploratory_test}
\end{figure}

The central limit theorem available for the order statistics allows
the construction of an hypothesis test. Since our analysis leads us
towards heavy-tailed distributions models, as a precaution, a
statistical test was performed to verify if a rather simpler model can
be fitted to the data. The normality assumption was considered as null
hypothesis for the test. The test was performed for the data in each
cell, by considering that the normal distribution parameters are given
by the empirical quantiles as expained previously. The $p-$values were
computed using a $\chi^2$ distribution. In this context, the local
normality assumption for the perturbation marks is globally
rejected. Figure~\ref{exploratory_test} shows the result of testing
the normality of the $z_{0.95}$ empirical quantile computed around the
Jupiter's orbit.

Indeed, there exist regions where the normality assumptions cannot be
rejected for the considered quantile. Still, the regions where this
hypothesis is rejected clearly indicate that normality cannot be
assumed entirely. Therefore, a parametric statistical model has to be
able to reflect this situation~: indicate whenever is the case how
``heavy'' or how stable are the distributions tails.

The only parameter used during this exploratory analysis was the
partitioning of the location domain $K$. There is one more question to
answer~: do the obtained results depend on the patterns exhibited by
the data, or they are just a consequence of the partitioning in cells
of the data locations~? To answer this question, a bootstrap procedure
and a permutation test were implemented~\citep{DAVetal:97}.

\begin{figure*}[!htbp]
\begin{center}
\begin{tabular}{cc}
a)\epsfxsize=5.5cm \epsffile{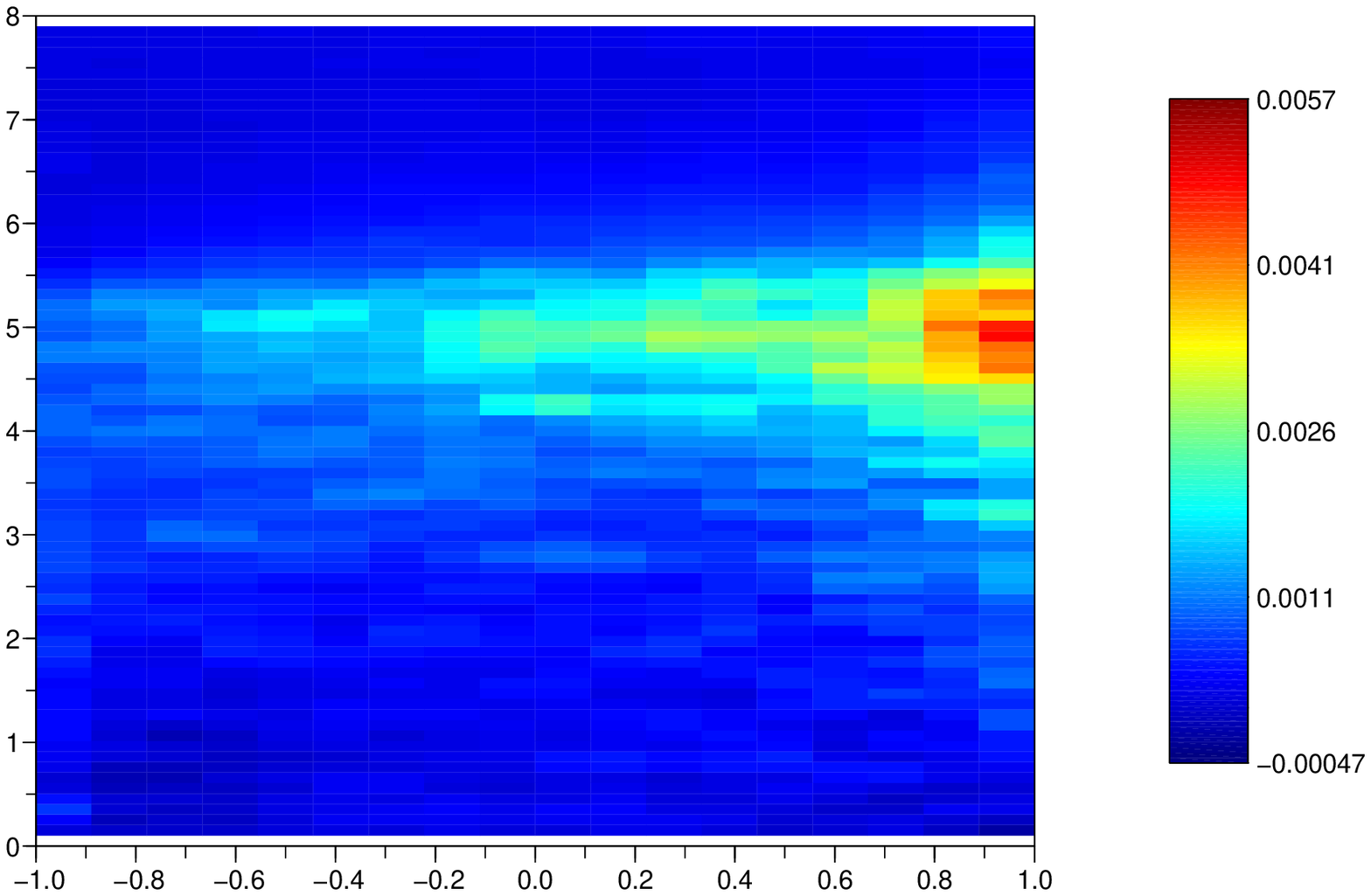} & 
b)\epsfxsize=5.5cm \epsffile{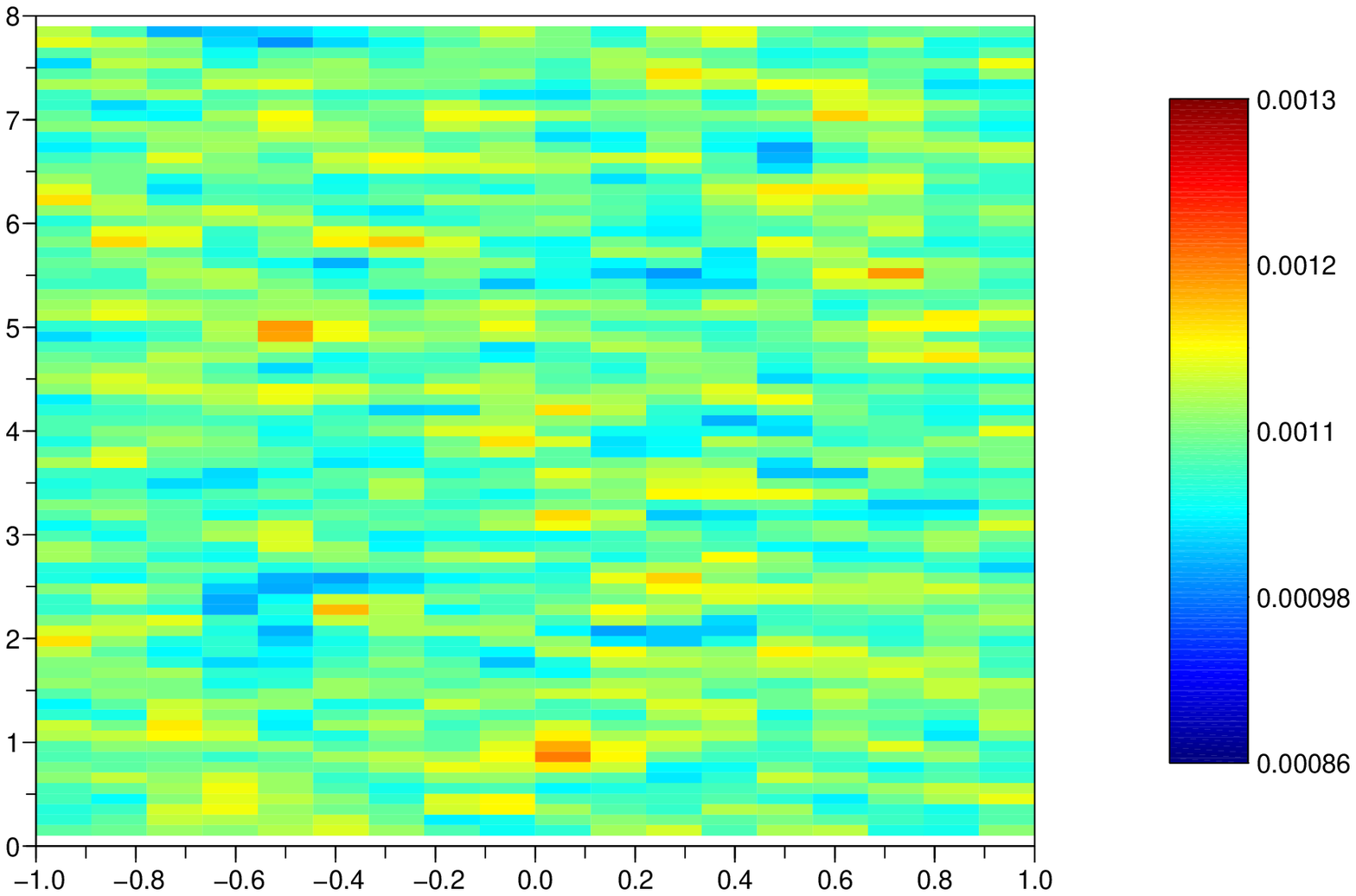}\\
\end{tabular}
\end{center}
\caption{Validation of the analysis based on the computation of
the difference indicator $z_{0.99} - \widehat{n}_{0.99}$ around Jupiter~: a)
bootstrap procedure ; b) permutation test.}
\label{exploratory_bootstrap}
\end{figure*}

Bootstrap samples were randomly selected by uniformly choosing $20\%$
from the entire perturbations data set. Difference indicators were
computed for this special data set. This operation was repeated $100$
times. At the end of the procedure, the empirical means of the
difference indicators were computed. In
Figure~\ref{exploratory_bootstrap}a the bootstrap mean of the
indicator $z_{0.99} - \widehat{n}_{0.99}$ around Jupiter's orbit is
showed. As expected, the same pattern is obtained as in
Figure~\ref{exploratory_tails}a~: important values are grouped around
the planet's orbit while exhibiting an arrow-like shape pointing from
right to left.

The permutation test follows the same steps as the bootstrap procedure
except that the perturbations are previously permuted. This means that
all the perturbations are modified as it follows~: for a given
perturbation, its mark is kept while its location is exchanged with
the location of another randomly chosen perturbation. This procedure
should destroy any pre-existing structure in the data. In this case,
we expect that applying a bootstrap procedure on this new data set
will indicate no relevant patterns. In
Figure~\ref{exploratory_bootstrap}b the result of such permutation
test is showed. The experiment was carried out in the vicinity of
Jupiter's orbit. After permuting the perturbations as indicated, the
previously described bootstrap procedure was applied in order to
estimate bootstrap means of the difference indicator $z_{0.99} -
\widehat{n}_{0.99}$. The result confirmed our expectations, in the
sense that no particular structure or pattern is observed. This
clearly indicates, that the analysis results were due mainly to the
original data structure and not to the partitioning of the
perturbations location domain in cells.

In the same time, the permutation test is also a verification of the
proposed exploratory methodology. This methodology depends on a
precision parameter for characterising the hidden structure or pattern
exhibited by the data. Still, whenever such a structure does not exist
at all, the present method detects nothing.

\subsection{Inference using heavy-tail distributions}
The empirical observations of the perturbations marks distributions
indicated fat tails and skewness behaviour. This leptokurtic character
of the perturbation distributions was observed especially in the
vicinity of the planets orbits. In response to this empirical evidence
heavy-tail distribution modelling was chosen.

The same cell partitioning as for the exploratory analysis is
maintained. The previously mentioned algorithm for estimating stable
laws parameters was run for the data in each cell.

\begin{figure*}[!htbp]
\begin{center}
\begin{tabular}{cc}
a)\epsfxsize=5.5cm \epsffile{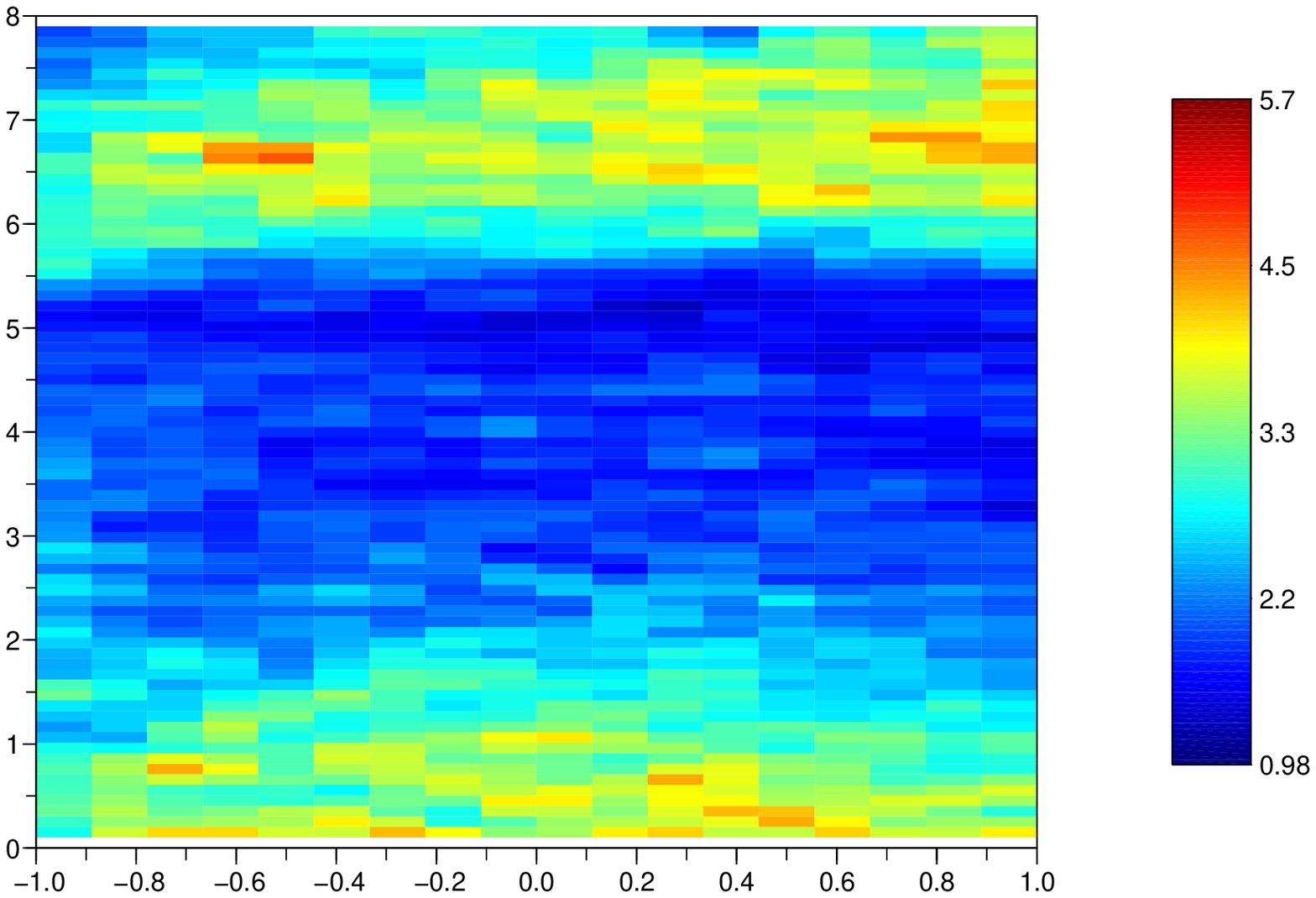} & 
b)\epsfxsize=5.5cm \epsffile{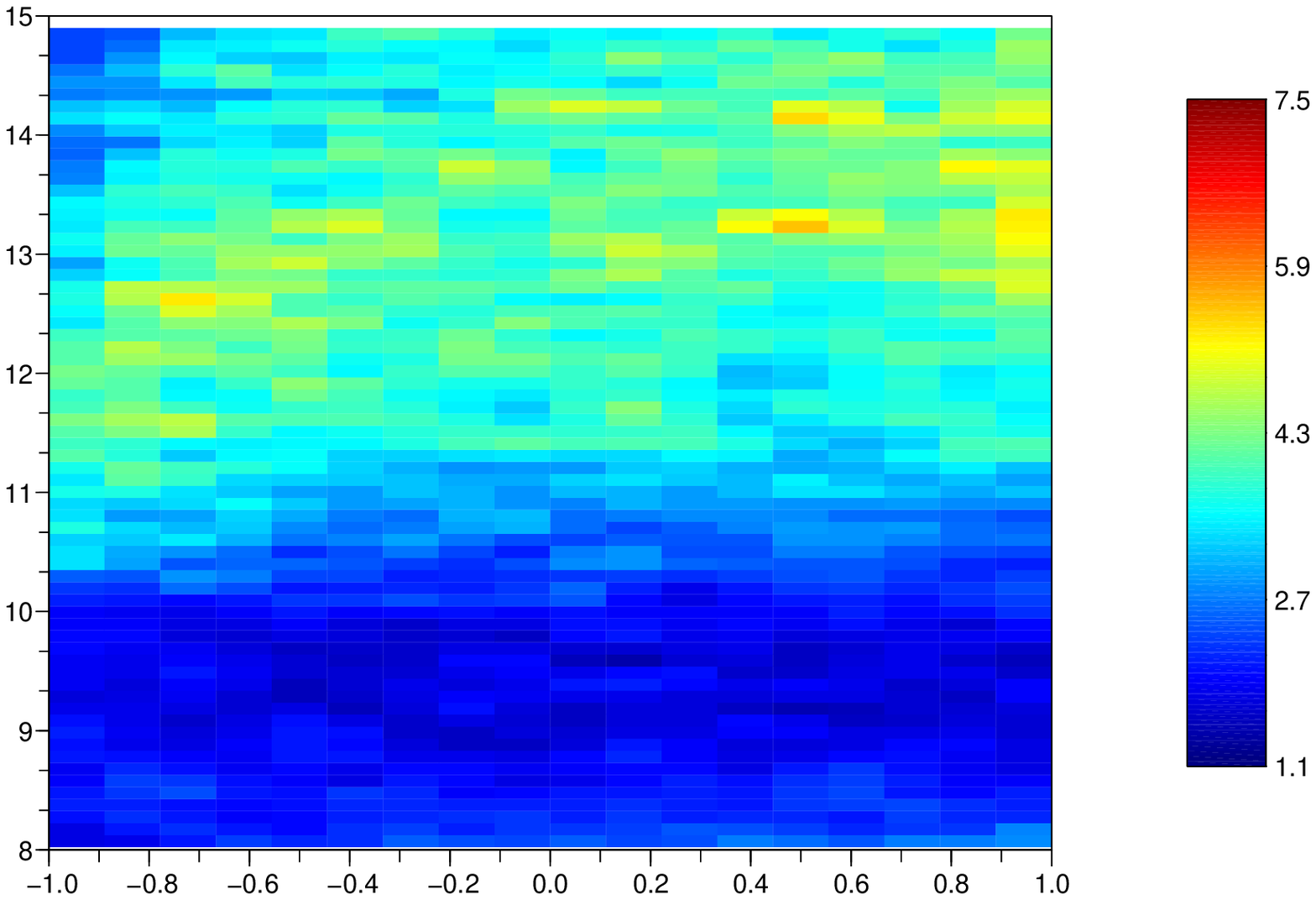}\\
c)\epsfxsize=5.5cm \epsffile{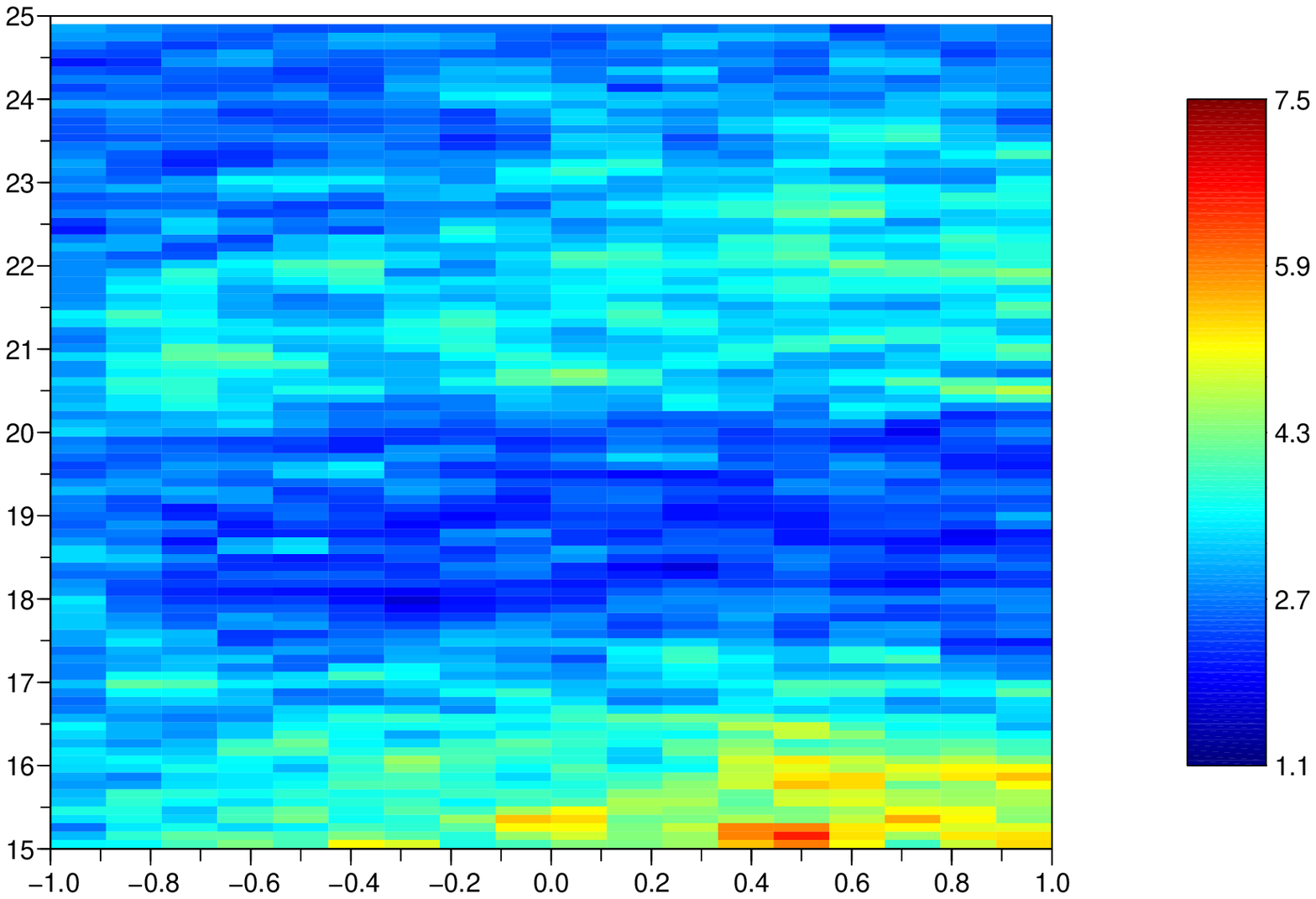} & 
d)\epsfxsize=5.5cm \epsffile{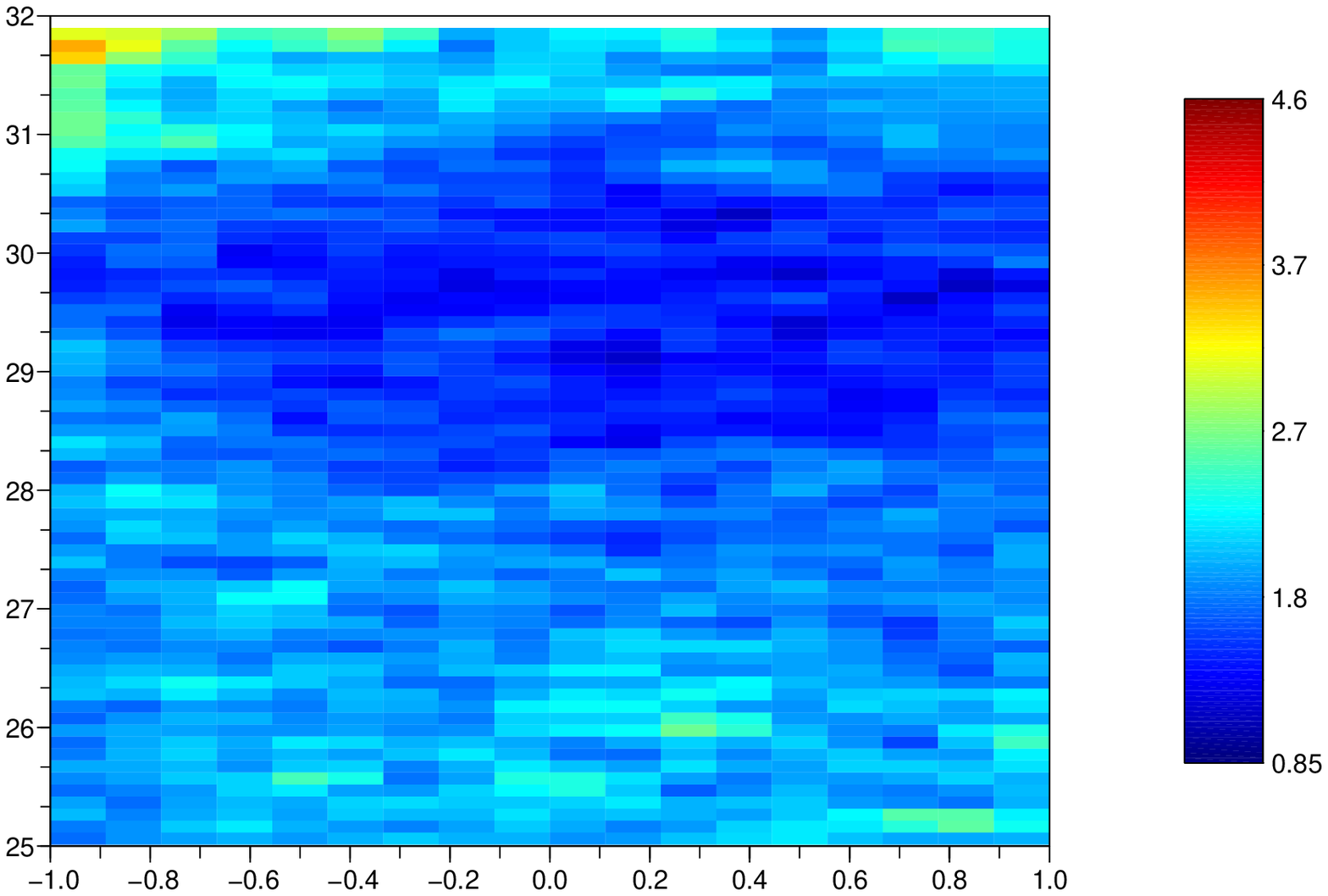}\\ 
\end{tabular}
\end{center}
\caption{Estimation result of the tail exponent $\alpha$ for the
perturbation marks around the big planets~: a) Jupiter, b) Saturn, c)
Uranus d) Neptune.}
\label{stable_alpha}
\end{figure*}

In Figure~\ref{stable_alpha} the estimation result of the tail
exponent is shown. Clearly, it can be observed a region formed by the
cells corresponding to estimated $\alpha$ values lower than $2$. This
kind of region may be located around each orbit corresponding to a big
planet. The shape of this region is less picked than the region
obtained using empirical quantiles. Still, the two results are
coherent. Both results indicate that the heavy-tailed character of the
perturbations distributions exhibits a spatial pattern. This spatial
pattern is located around the orbits of the big planets.

\begin{figure}[!htbp]
\begin{center}
\begin{tabular}{c}
\epsfxsize=5.5cm \epsffile{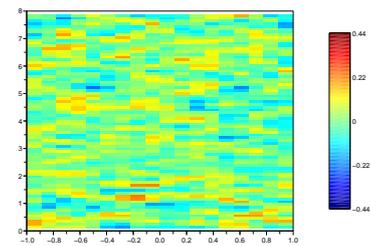}
\end{tabular}
\end{center}
\caption{Estimation result of the skewness parameter $\beta$ for the
perturbations marks around Jupiter.}
\label{stable_beta}
\end{figure}

The skewness of the data distribution can be analysed by looking at
the results shown in Figure~\ref{stable_beta}. Indeed, it can be
observed that there are cells containing perturbations following a
skewed distribution. The obtained results indicate neither the
presence of a pattern by such distributions, nor the presence of such
a pattern around the orbits of the big planets.

\begin{figure*}[!htbp]
\begin{center}
\begin{tabular}{cc}
a)\epsfxsize=5.5cm \epsffile{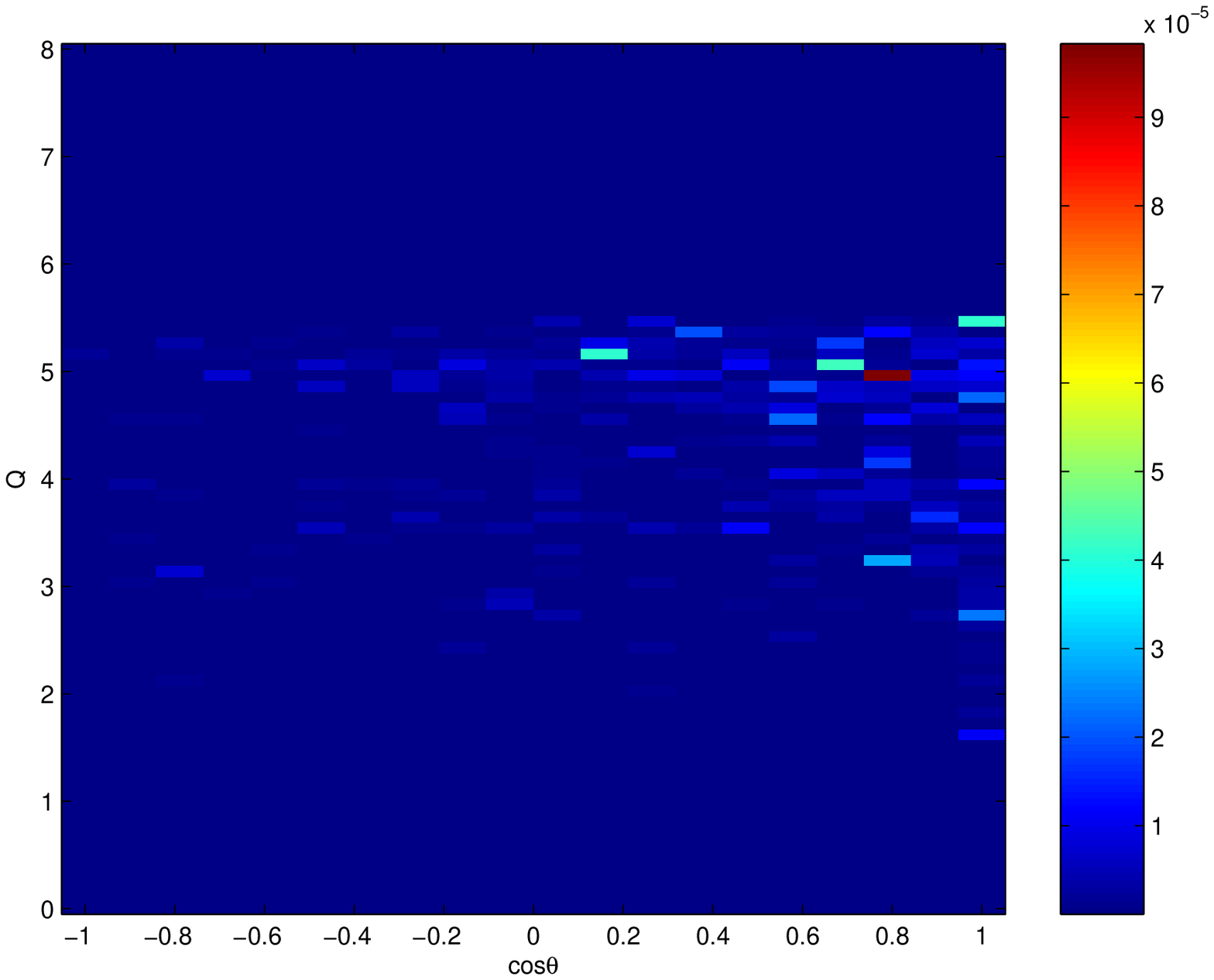} & 
b)\epsfxsize=5.5cm \epsffile{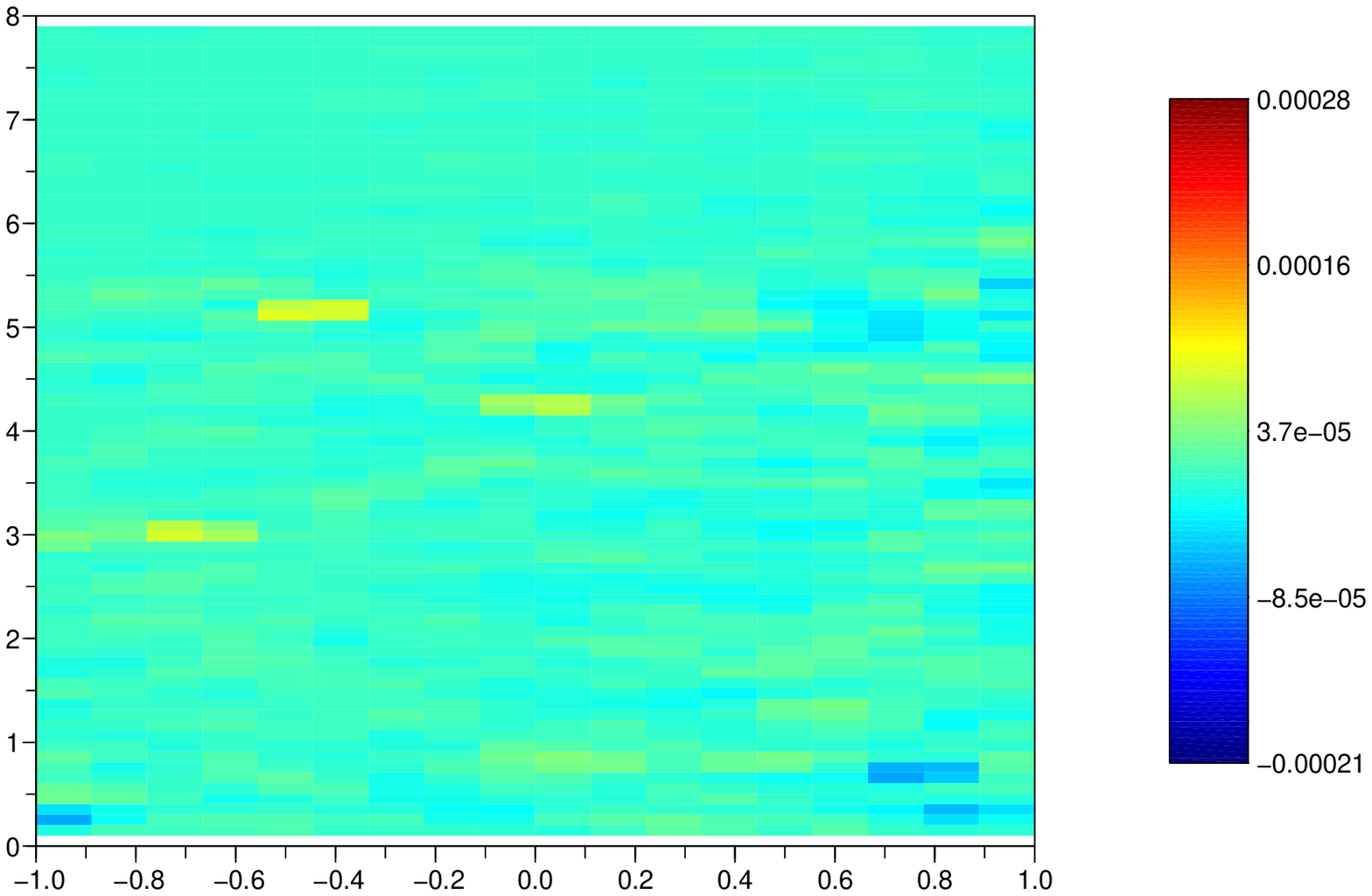}
\end{tabular}
\end{center}
\caption{Estimation result of the scale parameter $\sigma$ and shift
parameter $\delta$ for the perturbation marks around Jupiter.}
\label{stable_sigmadelta}
\end{figure*}

The estimation results for the $\sigma$ and $\delta$ parameters are
presented in Figure~\ref{stable_sigmadelta}. The scale parameter
indicates how heavy are the distribution tails. In
Figure~\ref{stable_sigmadelta}a, it may be observed that the most
important values of $\sigma$ tend to form a spatial pattern similar
with the patterns formed by the difference indicator based on order
statistics and the tail exponent, respectively. The results
obtained for the $\delta$ parameter indicate that a shift of the
perturbation may exist around the orbit of the corresponding big
planets.

\begin{figure}[!htbp] 
\begin{center} 
\begin{tabular}{c} 
\epsfxsize=5.5cm \epsffile{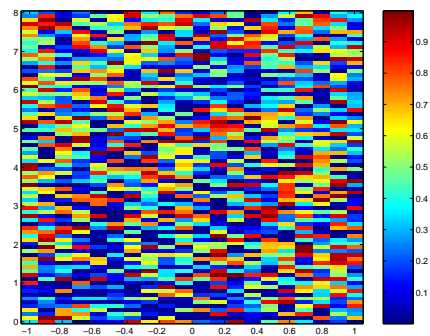} 
\end{tabular} 
\end{center}
\caption{$p-$values computed for testing if the empirical quantiles
$z_{0.99}$ around the Jupiter's orbit are originated from a heavy-tail
distribution.}  
\label{stable_test} 
\end{figure}

In order to check these results a statistical test using the central
limit theorem for order statistics was built. Clearly, this result can
be used in order to verify if the empirical quantiles from a cell are
coming rather from the distribution characterised by the parameters
previously estimated. Figure~\ref{stable_test} shows the result of a
test verifying that the $z_{0.99}$ quantiles around the Jupiter's
orbit are originated from a heavy-tail distribution, while the
quantiles outside this region are coming rather from Pareto
distribution. It can be observed that high values for the $p-$values
are spread around the entire region~: for $81.5\%$ of the cells we
cannot reject the null hypothesis. Clearly, this result shows a far
better characterisation of the distribution tails of the perturbations
than the test for the normality assumption performed in the preceding
section.

\begin{figure*}[!htbp]
\begin{center}
\begin{tabular}{cc}
a)\epsfxsize=5.5cm \epsffile{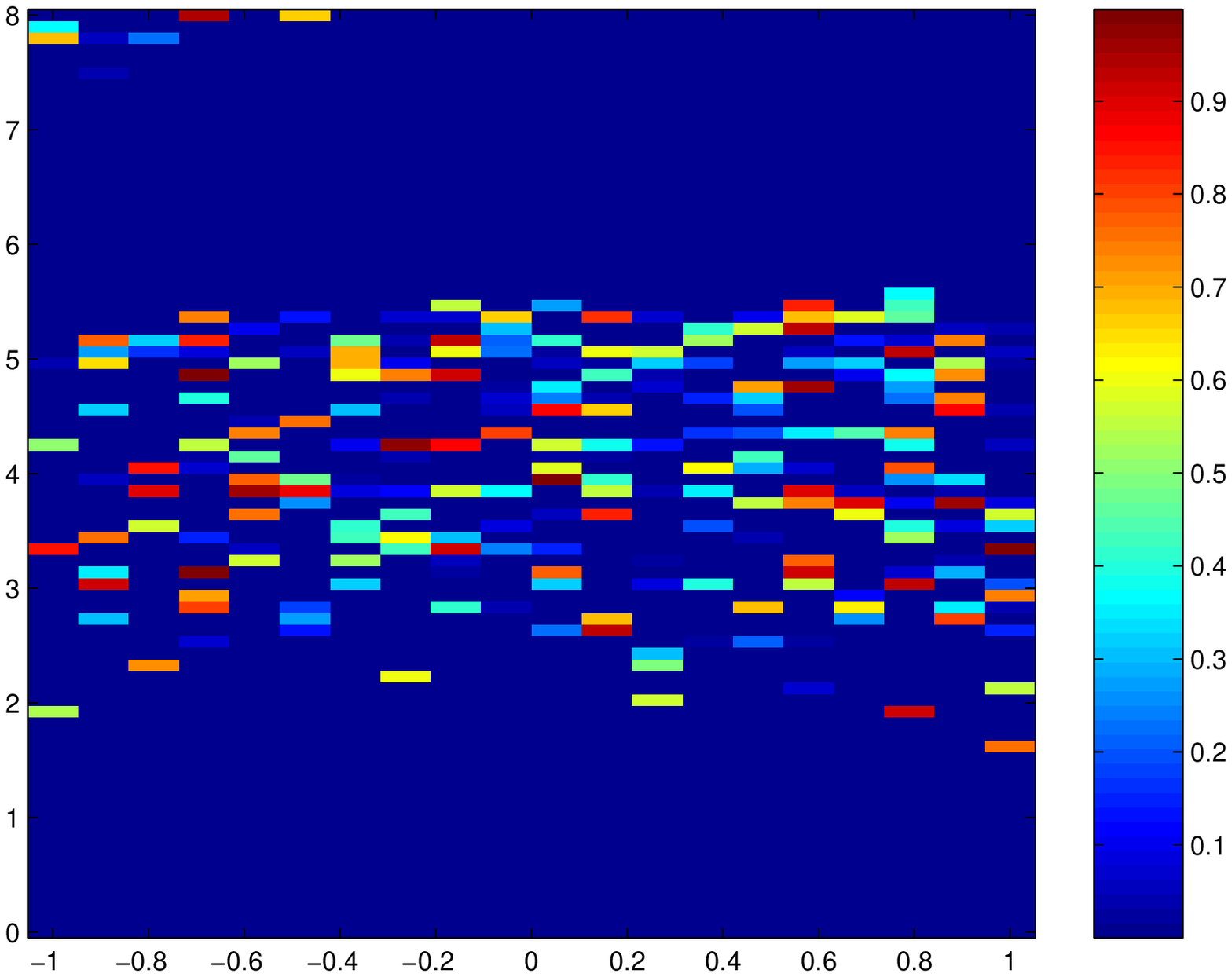} & 
b)\epsfxsize=5.5cm \epsffile{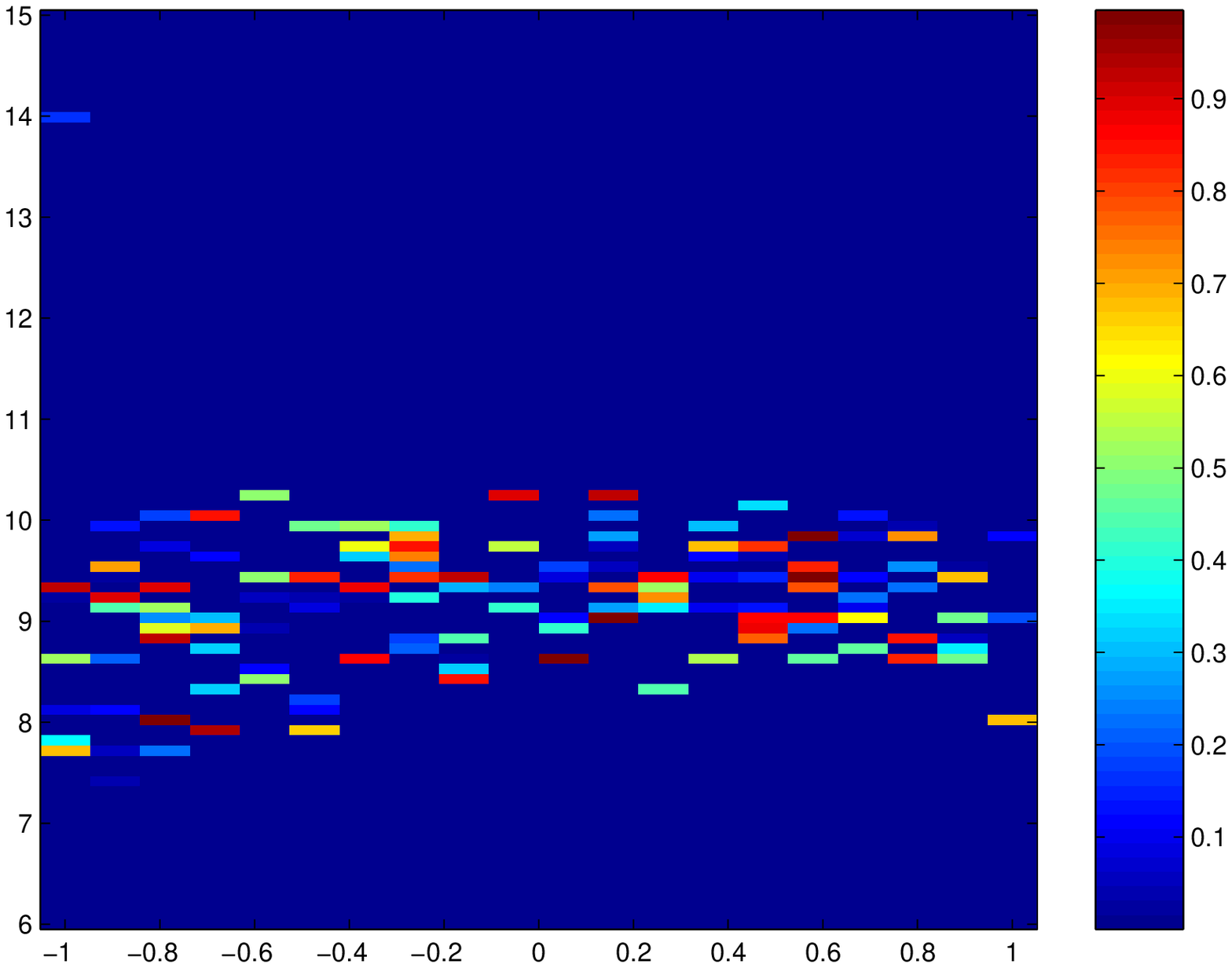}\\
c)\epsfxsize=5.5cm \epsffile{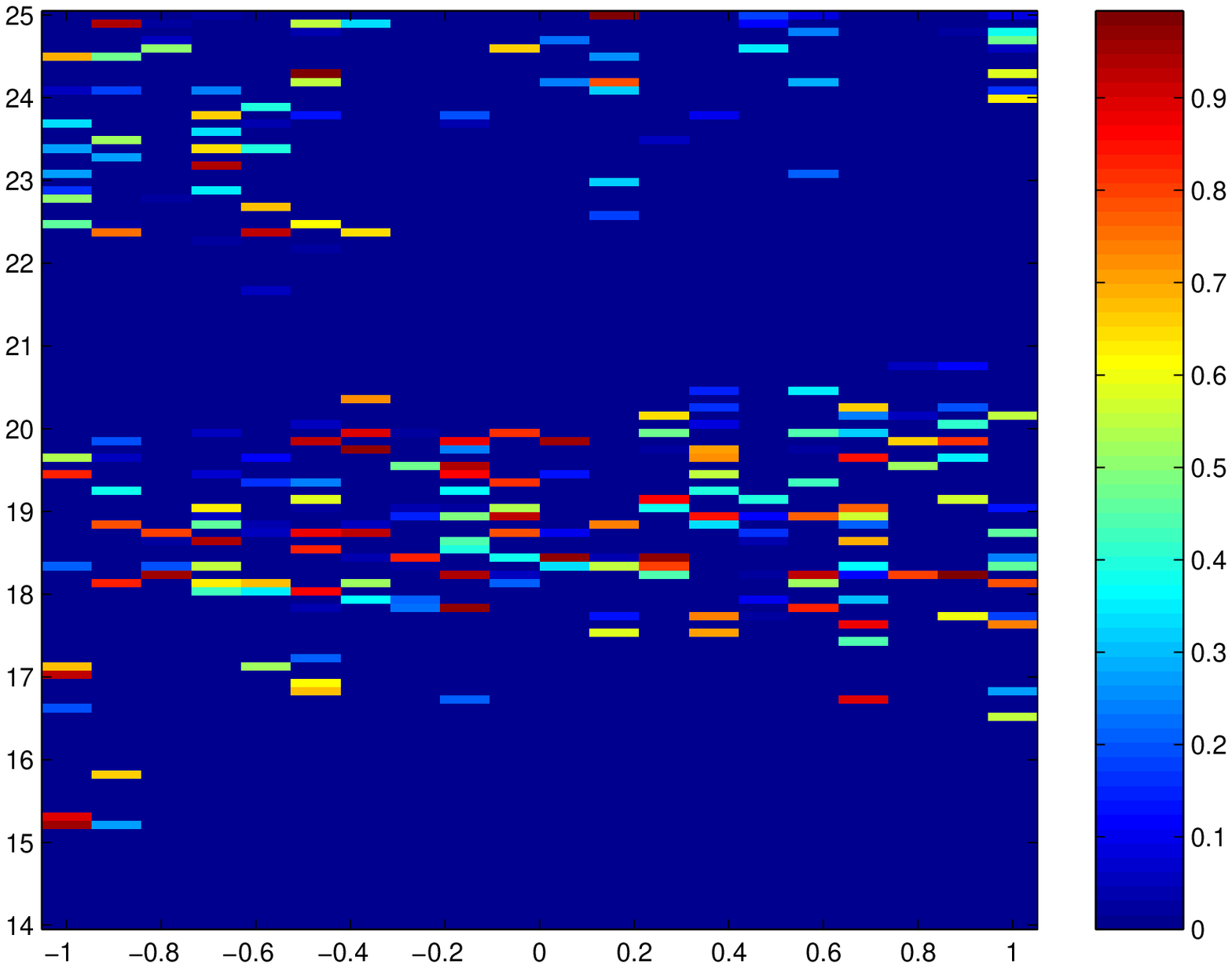} & 
d)\epsfxsize=5.5cm \epsffile{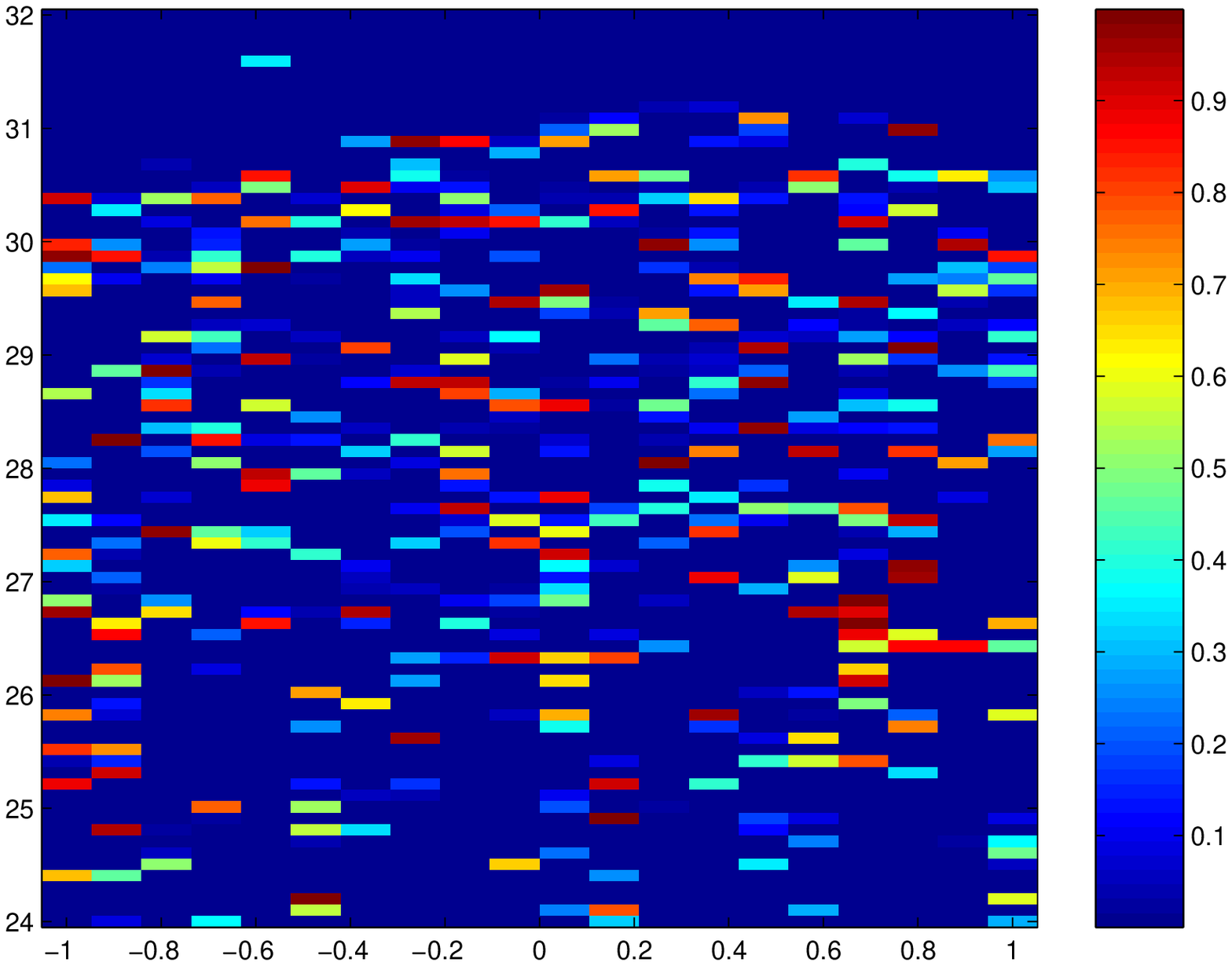}\\ 
\end{tabular}
\end{center}
\caption{$p-$values of a $\chi^2$ statistical test for the
perturbations with $\alpha < 2$ around the big planets~: a) Jupiter,
b) Saturn, c) Uranus, d) Neptune.}
\label{stable_alpha2}
\end{figure*}

The previous test certifies the perturbations distributions tails exhibit a
stable or regular variation behaviour. If the perturbations are close
to the orbit of a big planet then they have rather a stable
behaviour. Figure~\ref{stable_alpha2} shows the $p-$values of
a $\chi^{2}-$test implemented for the perturbations with estimated tail
exponent $\alpha < 2$. This test allows to check the perturbations
also for their distribution body. It can be observed that almost in
all these regions the assumption of stable distributions is not
rejected.

For the perturbations with a tail exponent greater than $2$, an
alternative family of distributions with regularly varying tails was
considered for modelling. Its expressions is given below~:
\begin{equation}
f(z) = \frac{C_{\kappa,\alpha}}{1+\mid \kappa z - \omega\mid^{\alpha+1}},
\label{regular_variation}
\end{equation}
with $C_{\kappa,\alpha}$ the normalising constant, $\kappa$ the scale
parameter, $\omega$ the location parameter and $\alpha$ the tail
exponent.

The parameter estimation for such distributions was done in several
steps. First, the tail exponent $\alpha$ was considered obtained from
the previous algorithm. Second, the location parameter $\omega$ was
estimated by the empirical mean of the data samples. Finally, the
normalising constant $C_{\kappa,\alpha}$ and the scale parameter
$\kappa$ were estimated using the method of moments.

\begin{figure*}[!htbp]
\begin{center}
\begin{tabular}{cc}
a)\epsfxsize=5.5cm \epsffile{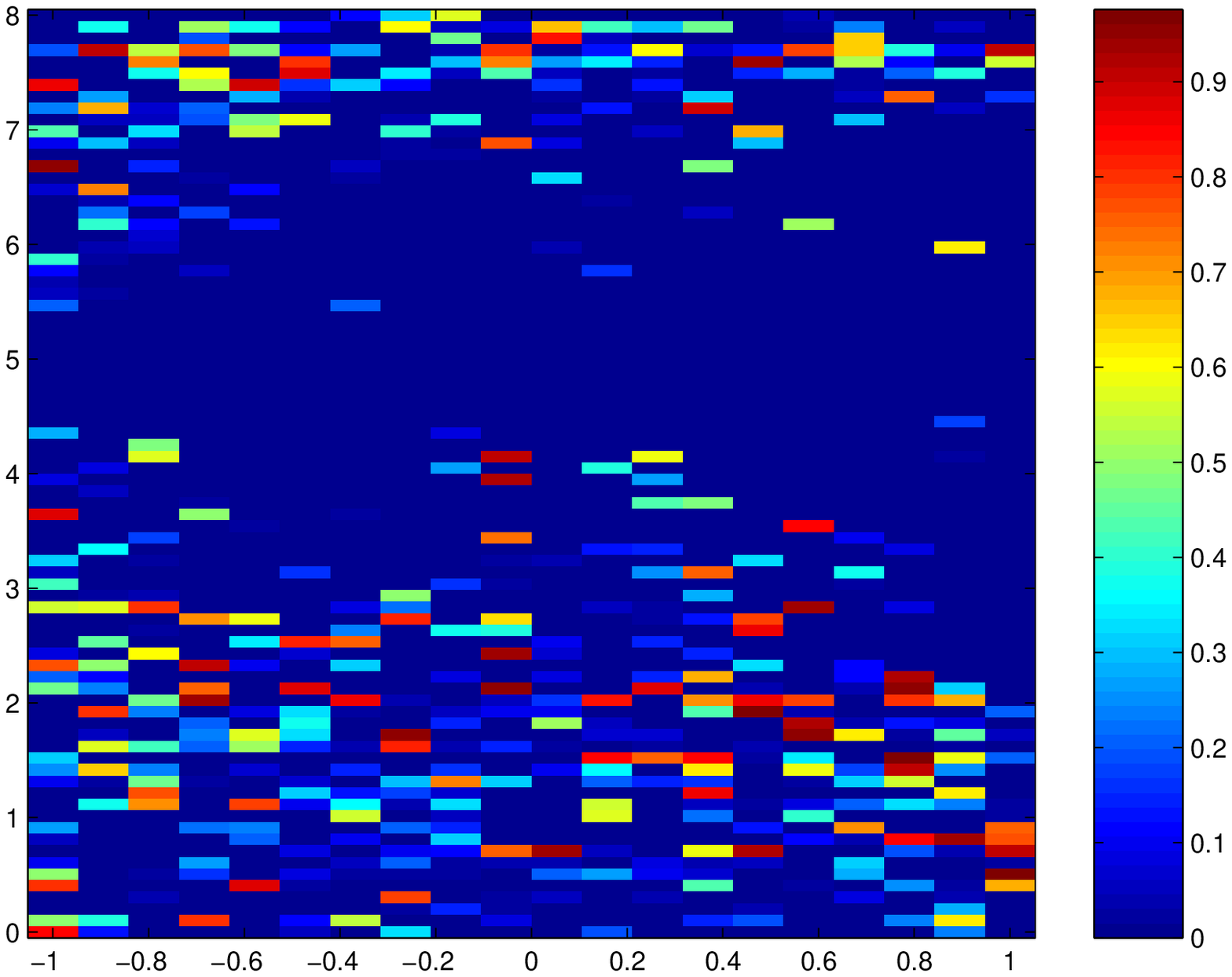} & 
b)\epsfxsize=5.5cm \epsffile{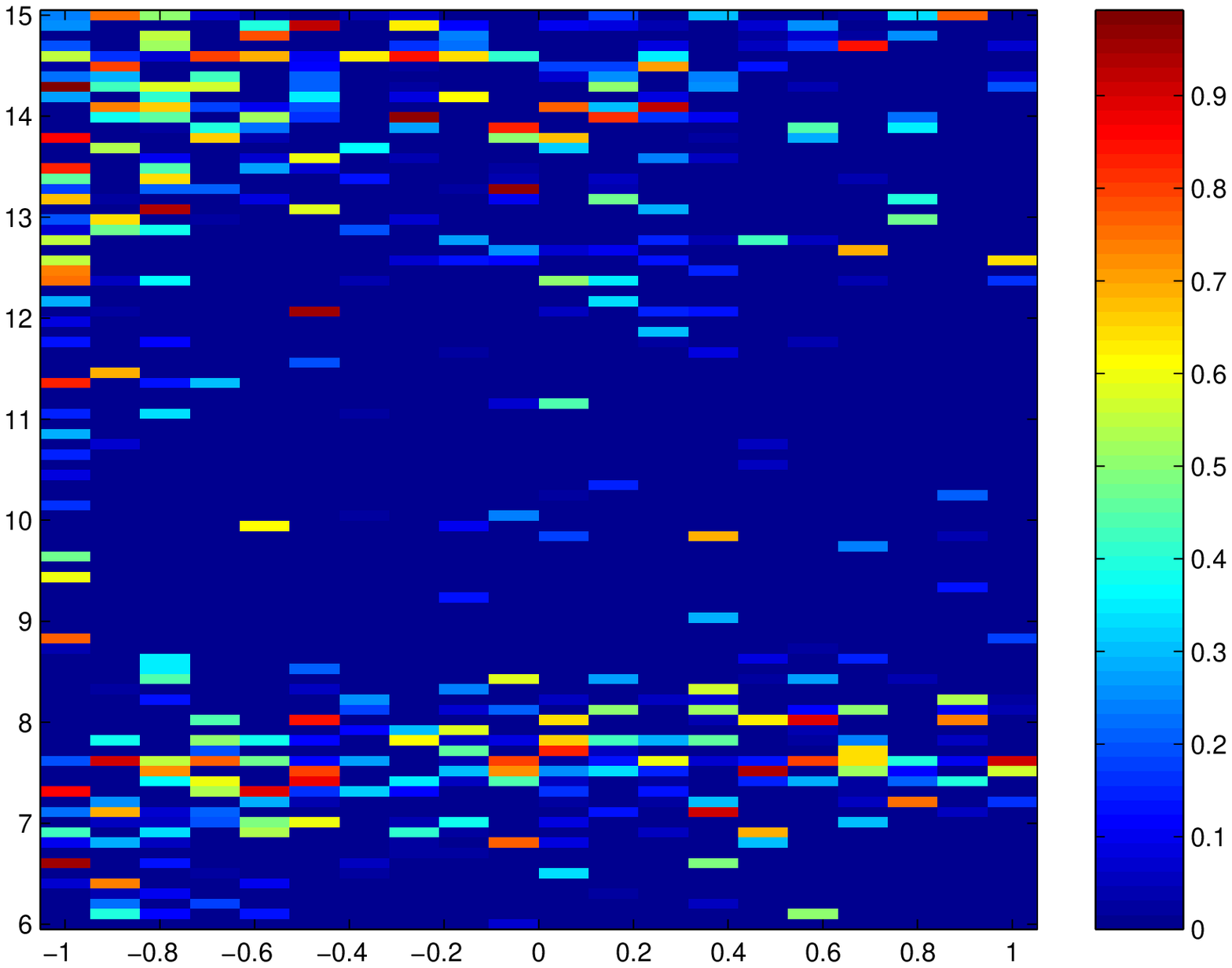}\\
c)\epsfxsize=5.5cm \epsffile{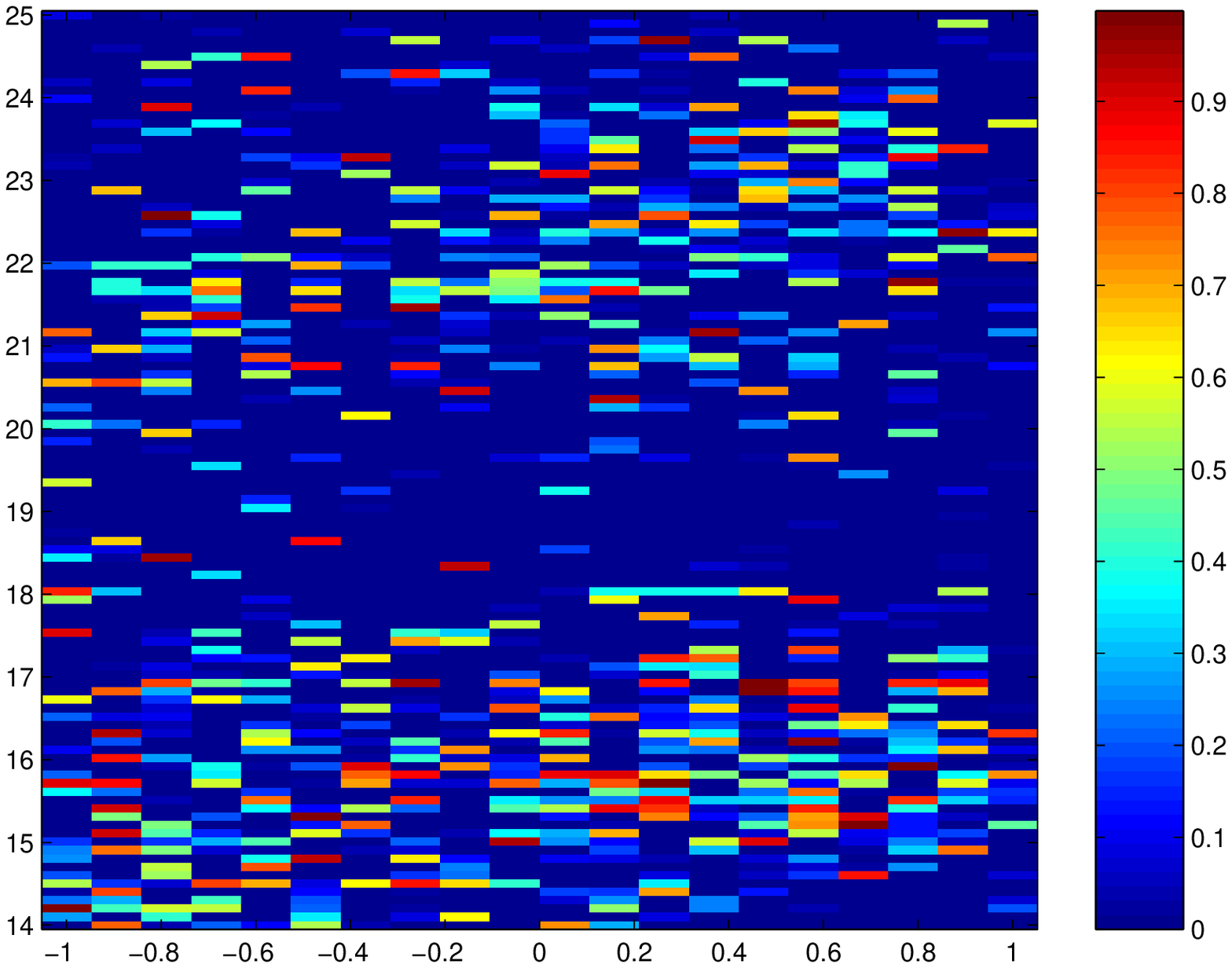} & 
d)\epsfxsize=5.5cm \epsffile{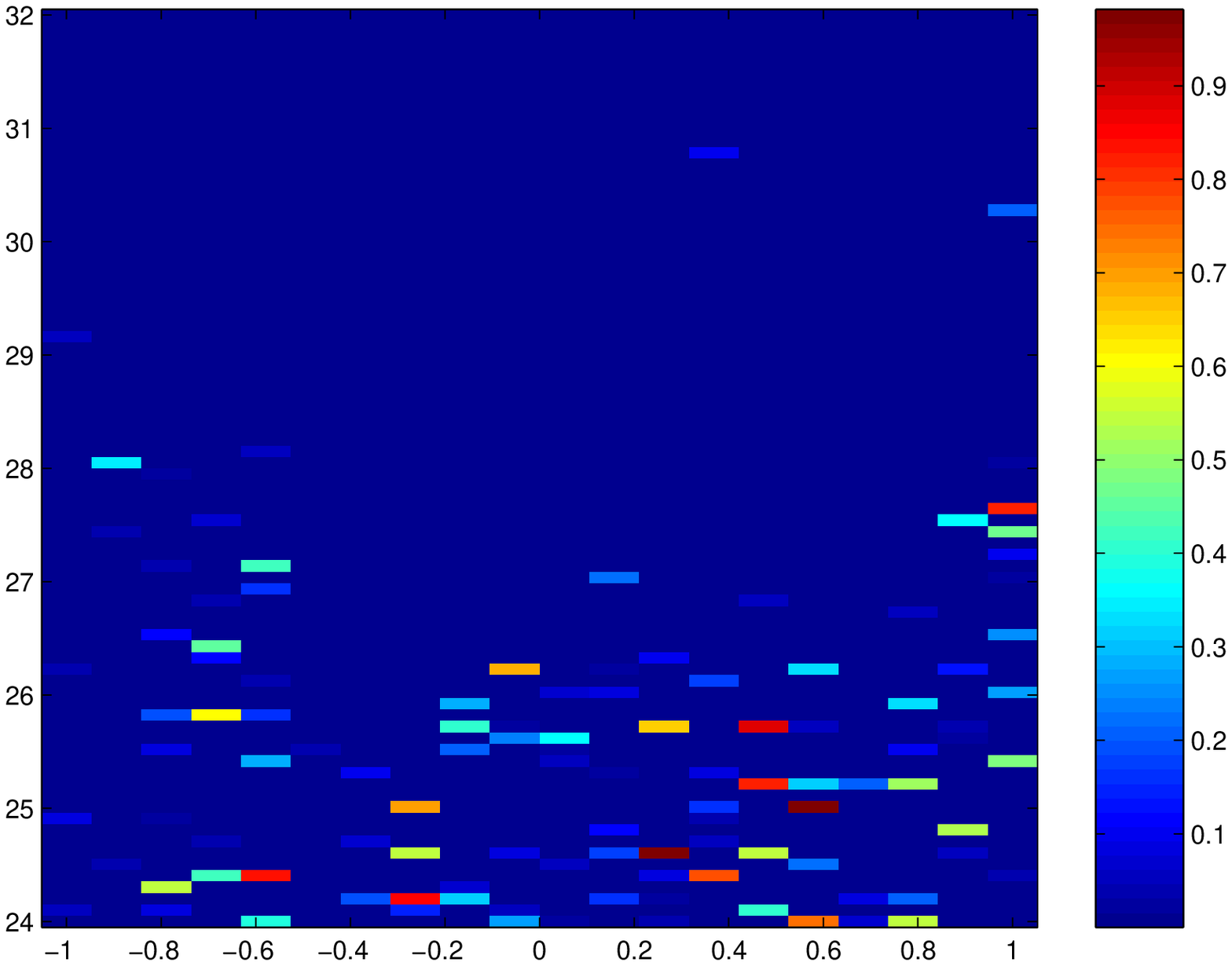}\\ 
\end{tabular}
\end{center}
\caption{$p-$values of a $\chi^2$ statistical test for the
perturbations with $\alpha \geq 2$ around the big planets~: a)
Jupiter, b) Saturn, c) Uranus, d) Neptune.}
\label{vr_nonstab}
\end{figure*}

A $\chi^2$ statistical test was done for the perturbations with
$\alpha \geq 2$. The null hypothesis considered was that the
considered perturbations follow a regularly varying tails
distribution~(\ref{regular_variation}) with parameters given by the
previously described procedure. The obtained $p-$values are shown in
Figure~\ref{vr_nonstab}. It can be noticed that in the majority of
considered cells the null hypothesis is not rejected.

\section{Discussion and interpretation}
Some of the features present in the Figures can be explained in the
framework of the analytical theory of close encounters \citep{OPI:76,
GREetal:88, CARetal:90, VAL.MAN:97}.

Let us consider the magnitude of the perturbations in the vicinity of
$a=a_{J}=5.2$~AU (Jupiter). The colour coding of the Figure
~\ref{exploratory_tails} represents the
magnitude $P$ of the perturbation, corresponding to
\begin{equation}
Z=-\frac{1}{a_{f}}+\frac{1}{a_{i}}\propto h_{f}-h_{i}  \label{eq:perturbation}
\end{equation}
where $a$ and $h$ are respectively the orbital semi-major axis and the
orbital energy of the heliocentric keplerian motion of the comet. The
subscripts $\,_{i}$ and $\,_{f}$ stand, respectively, for
\emph{initial} and \emph{final}, {\it i.e.}, before and after the
interaction with Jupiter).

Perturbations at planetary encounters are characterised by large and
in general asymmetric tails, as was shown by various authors
\citep{EVE:69,OIK.EVE:79, FRO.RIC:81}; an analytical explanation of
these features was given by \cite{CARetal:90} and by
\cite{VALetal:00}, and the consequences on the orbital evolution of
comets was discussed by \cite{VAL.MAN:97}.

Let us consider the case of parabolic initial orbits (our orbits are
in fact very close to parabolic). In the $q$-$\cos{i}$ plane, the
condition for the tails of the energy perturbation distribution to be
symmetric is:
\begin{eqnarray*}
0 & = & \frac{1-3+2\sqrt{2q/a_p}\cos{i}}{2\sqrt{3-2\sqrt{2q/a_p}\cos{i}}},
\end{eqnarray*}
where $a_p$ is the orbital semi major axis of the planet encountered.

However, the finite size of the available perturbation sample must be
taken into account, as the tails would become sufficiently populated
to show any asymmetry only for very large samples.

A way to take this effect into account is to consider that in
different regions of the $q$-$\cos{i}$ plane the probability $p$ for
the comet on a parabolic orbit to pass within a given unperturbed
distance $b$ from the planet would be, according to \cite{OPI:76}:
\begin{eqnarray*}
p 
& = & \frac{b^2}{a_p^2}\frac{\sqrt{3-2\sqrt{2q/a_p}\cos{i}}}{\pi\sin{i}\sqrt{2-2q/a_p}}.
\label{eq-perturbations}
\end{eqnarray*}

To take into account the size of the perturbation, we consider that
the angle $\gamma$ by which the planetocentric velocity of the comet
is rotated is given by:
\begin{eqnarray*}
\tan\frac{\gamma}{2} & = & \frac{a_pm_p}{bm_\odot\left(3-2\sqrt{2q/a_p}\cos{i}\right)};
\end{eqnarray*}
we then define a function $f$ as:
\begin{eqnarray}
f & = & p\tan\frac{\gamma}{2} \\
& = & \frac{bm_p}{a_pm_\odot\pi\sin{i}\sqrt{2-2q/a_p}\sqrt{3-2\sqrt{2q/a_p}\cos{i}}} \nonumber.
\end{eqnarray}

\begin{figure}[!htbp]
\begin{center}
\epsfxsize=7cm \epsffile{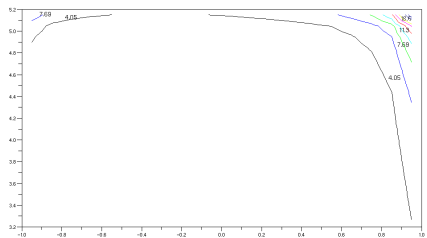}
\end{center}
\caption{Level curves for the function $f$ around the semi-major axis of Jupiter.}
\label{opik_theory}
\end{figure}

Figure~\ref{opik_theory} shows the level curves of $f$; as can be
seen, in it are reproduced the main features of
Figure~\ref{exploratory_tails}. The arrow-like
shape observed during the statistical study can be now observed on
the definition domain imposed by~(\ref{eq-perturbations}). 
This strenghten our interpretation of the features of Fig.~\ref{exploratory_tails} as
due to the geometry of close approaches described by \"Opik theory.

\section{Conclusion and perspectives}
\label{sec:conclusion}
In this paper a statistical study of the planetary perturbations on
Oort cloud comets was carried out. The exploratory analysis of the
perturbations distributions based on order statistics indicated the
tail behaviour as determinant feature. Following this idea, parametric
inference for heavy-tail distributions was implement. The obtained
results indicated that the perturbations following heavy-tail stable
distributions that are not always symmetric while tending to form a
spatial pattern. This pattern is rather arrow-like shaped and is
situated around the orbits of the big planets. A theoretical study was
carried out, and it was observed that this pattern is similar with the
theoretical curves derived from the \"Opik theory. The perturbations
outside this arrow shaped region were not exhibiting a stable
character and they were modelled by a family of distributions with
regularly varying tails. In both cases, stable and non-stable
distributions, the modelling choices were confirmed by a statistical
test.

Clearly, these choices and the estimation parameter estimation
procedures can be further improved. Nevertheless, the obtained results
give good indications and also good reasons for developing a
probabilistic methodology able to simulate such planetary
perturbations.

\begin{acknowledgements}
The authors are grateful to Alain Noullez for very useful comments
and remarks. Part of this research was supported by the University of
Lille 1 through the financial BQR program.
\end{acknowledgements}

\bibliographystyle{aa}
\bibliography{biblio}

\end{document}